\documentstyle[psfig,referee,epsf,color]{mn}

\newif\ifAMStwofonts

\ifoldfss
  \ifCUPmtlplainloaded \else
    \NewTextAlphabet{textbfit} {cmbxti10} {}
    \NewTextAlphabet{textbfss} {cmssbx10} {}
    \NewMathAlphabet{mathbfit} {cmbxti10} {} 
    \NewMathAlphabet{mathbfss} {cmssbx10} {} 
  \fi
  \ifAMStwofonts
    \ifCUPmtlplainloaded \else
      \NewSymbolFont{upmath} {eurm10}
      \NewSymbolFont{AMSa} {msam10}
      \NewMathSymbol{\upi}     {0}{upmath}{19}
      \NewMathSymbol{\umu}     {0}{upmath}{16}
      \NewMathSymbol{\upartial}{0}{upmath}{40}
      \NewMathSymbol{\leqslant}{3}{AMSa}{36}
      \NewMathSymbol{\geqslant}{3}{AMSa}{3E}

       \let\le=\leqslant
      \let\geq=\geqslant \let\ge=\geqslant
    \fi
  \fi
\fi 

\ifnfssone
  \newmathalphabet{\mathit}
  \addtoversion{normal}{\mathit}{cmr}{m}{it}
  \addtoversion{bold}{\mathit}{cmr}{bx}{it}
  \newmathalphabet{\mathbfit} 
  \addtoversion{normal}{\mathbfit}{cmr}{bx}{it}
  \addtoversion{bold}{\mathbfit}{cmr}{bx}{it}
  \newmathalphabet{\mathbfss} 
  \addtoversion{normal}{\mathbfss}{cmss}{bx}{n}
  \addtoversion{bold}{\mathbfss}{cmss}{bx}{n}
  \ifAMStwofonts
    \ifCUPmtlplainloaded \else
      \UseAMStwoboldmath
      \makeatletter
      \new@mathgroup\upmath@group
      \define@mathgroup\mv@normal\upmath@group{eur}{m}{n}
      \define@mathgroup\mv@bold\upmath@group{eur}{b}{n}
      \edef\UPM{\hexnumber\upmath@group}
      \new@mathgroup\amsa@group
      \define@mathgroup\mv@normal\amsa@group{msa}{m}{n}
      \define@mathgroup\mv@bold\amsa@group{msa}{m}{n}
      \edef\AMSa{\hexnumber\amsa@group}
      \makeatother
      \mathchardef\upi="0\UPM19
      \mathchardef\umu="0\UPM16
      \mathchardef\upartial="0\UPM40
      \mathchardef\leqslant="3\AMSa36
      \mathchardef\geqslant="3\AMSa3E

       \let\le=\leqslant
      \let\geq=\geqslant \let\ge=\geqslant
    \fi
  \fi
\fi 

\ifnfsstwo
  \DeclareMathAlphabet{\mathbfit}{OT1}{cmr}{bx}{it}
  \SetMathAlphabet\mathbfit{bold}{OT1}{cmr}{bx}{it}
  \DeclareMathAlphabet{\mathbfss}{OT1}{cmss}{bx}{n}
  \SetMathAlphabet\mathbfss{bold}{OT1}{cmss}{bx}{n}
  \ifAMStwofonts
    \ifCUPmtlplainloaded \else
      \DeclareSymbolFont{UPM}{U}{eur}{m}{n}
      \SetSymbolFont{UPM}{bold}{U}{eur}{b}{n}
      \DeclareSymbolFont{AMSa}{U}{msa}{m}{n}
      \DeclareMathSymbol{\upi}{0}{UPM}{"19}
      \DeclareMathSymbol{\umu}{0}{UPM}{"16}
      \DeclareMathSymbol{\upartial}{0}{UPM}{"40}
      \DeclareMathSymbol{\leqslant}{3}{AMSa}{"36}
      \DeclareMathSymbol{\geqslant}{3}{AMSa}{"3E}

       \let\le=\leqslant
      \let\geq=\geqslant \let\ge=\geqslant
    \fi
  \fi
\fi 

\ifCUPmtlplainloaded \else
  \ifAMStwofonts \else 
    \def\upi{\pi}
    \def\umu{\mu}
    \def\upartial{\partial}
  \fi
\fi

\newcommand{\be}{\begin{equation}}
\newcommand{\ee}{\end{equation}}
\def\bea{\begin{eqnarray}}
\def\eea{\end{eqnarray}}
\newcommand\eps{\varepsilon_{SN}}  
  
\newcommand{\epg}{\varepsilon_{g}}  
\newcommand{\msu}{M_{\odot}}  
\newcommand{\Af}{{$A_{Fe}$}}
\newcommand{\La}{{$\Lambda$CDM}}
\newcommand{\gr}{\kern 2pt\hbox{}^\circ{\kern -2pt K}} 
\newcommand{\oml}{\Omega_{\Lambda}}
\newcommand{\omm}{\Omega_{m}}
\newcommand{\brr}{\begin{array}}
\newcommand{\err}{\end{array}}
\newcommand{\ltsima}{$\; \buildrel < \over \sim \;$}
\newcommand{\simlt}{\lower.5ex\hbox{\ltsima}}
\newcommand{\gtsima}{$\; \buildrel > \over \sim \;$}
\newcommand{\simgt}{\lower.5ex\hbox{\gtsima}}

\title[Metal enrichment: SPH simulations]
{Iron abundances and heating of the ICM in hydrodynamical simulations of
galaxy clusters}

\author[R. Valdarnini]
{R. Valdarnini$^{1}$ 
\\
$^{1}$SISSA, Via Beirut 2-4 34014 Trieste Italy.}

\date{Accepted .....
      Received .....;
      in original form .....}

\pubyear{2002}

\begin{document}
\maketitle

\label{firstpage}

\begin{abstract}
Results from a large set of hydrodynamical  smoothed-particle-hydrodynamic
(SPH) simulations of galaxy clusters in a flat \La~cosmology are used to
investigate the metal enrichment and heating of the intracluster medium (ICM).
The physical modeling of the gas includes radiative cooling, star formation,
energy feedback and metal enrichment that follow from the explosions 
 of supernovae of type II and Ia.
The metallicity dependence of the cooling function is also taken into account.
The gas is metal enriched from star particles according to the SPH 
prescriptions. 
The simulations have been performed to study the dependence of 
final metal abundances and heating of the ICM on the numerical
 resolution and the model parameters.

For a fiducial set of model  prescriptions the results indicate 
radial iron profiles in broad agreement with observations; global
iron abundances are also consistent with data.
It is found that the iron distribution in the intracluster medium 
 is critically 
dependent on the shape of the metal deposition profile.
At large radii the radial iron abundance profiles in the simulations
are steeper than those in the data, suggesting a dynamical evolution 
of simulated clusters different from those observed.
For low temperature clusters simulations yield iron abundances below
the allowed observational range, unless it is introduced a minimum
diffusion length of metals in the ICM.
The simulated emission-weighted radial temperature profiles are in good
agreement with data for cooling flow clusters, but 
 at very small distances from the cluster centres ($\sim 2\%$ of the 
virial radii) the temperatures are a factor $\sim$ two higher than
the measured spectral values.

The luminosity-temperature relation is in excellent agreement with the data,
cool clusters ($T_X\sim 1keV$) have a core excess entropy of $\sim 200 keVcm^2$
and their X-ray properties are unaffected by the amount of feedback
energy that has heated the ICM.
The findings support the model proposed recently by Bryan, where
the cluster X-ray properties are determined by radiative cooling.
The fraction of hot gas $f_g$ at the virial radius increases with
 $T_X$ and the distribution obtained from the simulated cluster sample 
is consistent with the observational ranges.

\end{abstract}


\section{INTRODUCTION}

Galaxy clusters are the largest virialized structures known in the
universe and are considered useful probes to constrain current cosmological
theories of structure formation.
X-ray observations of galaxy clusters show that most of the baryonic cluster
mass is in the form of a hot ionized intracluster medium at temperatures
$\simeq 10^7 -10^8~\gr$, with the bulk of the emission in the X-ray band from
bremsstrahlung processes \cite{sar86}.
The dependence of the X-ray emission on the square of the gas  density allows
 us to 
construct cluster samples without the biases which may arise in the optical
band.
The final physical state of the ICM is primarily 
determined by the gravitational processes, which have driven the dynamical
evolution of the gas and dark matter mass components during the cluster
collapse.
 Under the assumption of hydrostatic equilibrium the ICM gas distribution
can be modeled to connect the gas temperature $T_X$ to the cluster 
virial mass or to the bolometric X-ray luminosity $L_X$. The X-ray
temperature and luminosity functions are then predicted for a
given cosmological model according to the standard theoretical
Press-Schecter (1974) mass function.
The observed evolutionary history of these functions \cite{edg90,hen91,hen97} 
can then be used to put
severe constraints on the allowed cosmological models 
\cite{hen91,whi93,eke96,bah98,kit98}.

If the ICM has been shock heated solely by gravitational processes, the 
cluster scaling relations are predicted to obey a self-similar
behavior. For instance, the $L_X-T_X$ relation should scale 
as $L_X \propto T_X^2$ \cite{kai86}.
For clusters with $T_X \simgt 2keV$ there is a wide observational evidence
\cite{dav93,ala98,mar98}   
that the observed bolometric X-ray luminosity scales with
temperature with a slope steeper than expected ($L_X \propto T_X^3$).
 This implies that low temperature clusters have central densities lower
than those predicted by the self-similar scaling relations 
\cite{bow97,pon99,lly20}.
 This break of self-similarity is usually taken as a strong evidence that
non-gravitational heating of the ICM has played an important role in the ICM
evolution \cite{evr91,kai91,whi91}, at least for cool clusters.

The most considered source of energy which has been considered as a heating 
mechanism for the ICM is supernovae (SNe) driven-winds, which inject energy 
into the ICM through SNe explosions of type Ia and II \cite{whi91,low96}.
The energy input required to heat the gas at the entropy level necessary 
to reproduce the observed departure from self-similarity is estimated to
lie in the range $\simeq0.5-3keV$ per particle \cite{bal99,wu20,to01}.
Several authors have argued that it is unlikely that SNe can provide 
the required energy to heat the gas and have suggested active galactic 
nuclei \cite{va99,wu20}, as the extra energy source for ICM heating.
 Bryan (2000) has proposed the alternative view that radiative 
cooling and the subsequent galaxy formation \cite{per20},
 can explain the observed 
$L_X-T_X$ relation because of the removal of low-entropy gas at the 
cluster cores.

The estimates of the energy injected from SNe into the ICM are based on the 
measured abundances of metals in the ICM. 
The elements that have enriched the ICM have been 
synthesized in the SNe explosions of the 
 stellar population of the cluster.
Two enrichment mechanisms have been proposed: supernova-driven galactic winds 
from early-type galaxies \cite{mu97}, or ram pressure stripping of the enriched 
gas from galaxies \cite{gu72}.
Another possibility is that of a significant contribution to the metal
enrichment and heating of the ICM, associated with an early-epoch generation
of massive Population III stars \cite{lo01}.

 A large set of observations confirms that the ICM of galaxy clusters is 
rich in metals \cite{ar92,mu97,fu98,dua20,fin20,fu20,ma20}.
Analyses of X-ray spectra show that the abundance of heavy elements in the ICM
is nearly $\simeq1/3$ solar. These measurements provide a strong support for the SN scenario as a heating source for the ICM.

Abundance gradients have also been measured 
\cite{ez97,dub20,whi20,fin20,de01,ir01}.
For instance Ezawa et al. (1997) found a decline in the iron abundance 
of AWM7 from $\simeq0.5$ solar in the centre to $\simeq0.2$ solar at 
a distance of $\simeq500 Kpc$.
Measurements of the relative abundance of the heavy elements 
can be used to constrain the enrichment mechanism of the ICM and the 
energy input from SNe. Analysis of the spatial distribution of metallicity
gradients is also important to discriminate among the proposed
enrichment scenarios \cite{dub20}.
 From the observed metallicities one can estimate the SN energy released
 given a shape for the initial-mass-function (IMF) and a nucleosynthesis
yield model for the SN explosions.
Finoguenov, Arnaud \& David  (2001) have 
obtained the Si and Fe abundances using the X-ray data of a selected sample 
of X-ray clusters. From the measured abundances, 
they find a significant contribution to the energy per particle associated 
with SNe explosions at temperatures $\simeq3keV$.

These kinds of estimates of the SN energy input suffer from the theoretical
uncertainties in the yield models and in the form of the IMF \cite{gib97}.
Moreover, these estimates can  vary widely with
 the assumed spatial distribution of the ICM metals and the
transfer efficiency of the kinetic energy released in a SN explosion to the 
ambient gas \cite{kr00}.
Hydrodynamical simulations of cluster evolution have the advantage over
analytical methods that they take into account the dynamical evolution of 
the gas.
From the results of numerical simulations of galaxy formation, Kravtsov \&
Yepes (2000) conclude that it is unlikely
that SNe can provide the required energy input, 
 even assuming the existence of radial metallicity gradients 
in the ICM.

In order to implement self-consistently the metal enrichment of the ICM 
in hydrodynamical simulations of galaxy clusters, one first must consider
the effects of non-gravitational processes on the cluster gas 
distribution.
The effects of radiative cooling in simulated clusters have been 
investigated by a number of authors \cite{ann96,lew20,per20,yos00,val02}.
One of the main conclusions of these simulations is that the modeling of
radiative processes for the gas cannot be decoupled from a prescription
for turning cold, dense gas into stars. This is done in order to avoid 
unphysically high densities (cooling catastrophe).
The luminosities $L_X$ of the simulated clusters are found to be physically
plausible, provided that a suitable prescription for the treatment of the cold
 gas has been taken into account.

In a previous paper \cite{val02} results from a set of hydrodynamical SPH 
simulations of galaxy clusters have been used to investigate how final 
X-ray properties of the simulated clusters depend upon the simulation numerical
 resolution and the chosen star formation (SF) prescription. 
For a chosen SF model, final X-ray luminosities have been found to be 
numerically 
stable and consistent with data, with uncertainties of a factor $\simeq2$.
In these simulations the metal enrichment of the ICM has been included 
with a minimal number of prescriptions and it was shown that 
for the simulated clusters, the final iron abundances of the ICM are in
broad agreement with measured values.

Chemical evolution in hydrodynamical SPH simulations has already been
considered in a variety of contexts 
\cite{mu94,ra96,ca98,bu00,mo01,ed01,li02,ag01}.
Metzler \& Evrard (1994) have investigated 
 the metal enrichment of the ICM in P3MSPH simulations of galaxy clusters 
in a standard cold dark matter (CDM) scenario.
The authors have not included radiative 
cooling in the simulations and have adopted a phenomenological prescription
 to model the chemical enrichment.

In a more recent paper, Kravtsov \& Yepes (2000) have estimated SN heating of 
the ICM using 
fixed-grid Eulerian hydrodynamical simulations. However, the numerical
resolution of the simulations was not adequate to study the evolution of 
simulated clusters. The number of SNe occurring in a given cluster
was estimated statistically from many small-box galaxy formation simulations.
The implementation of a self-consistent metal enrichment model
for the ICM in hydrodynamical simulations is therefore important for 
investigations of
 the ICM metal evolution.
This is particularly relevant in connection with the observed metallicity 
gradients and for assessing the reliability of the calculated amount of ICM 
heating from SNe, inferred from measured metallicity abundances.

The main aim of this paper is to analyze, in hydrodynamical SPH
simulations of galaxy clusters, the dependence of the final iron abundance 
 on a number of model parameters that control the ICM metal enrichment.
This is done in order to obtain for the simulated clusters a final ICM 
distribution
which can consistently fit a set of observational constraints, such as the 
observed iron abundances and at the same time the luminosity-temperature 
relation.
Implications for the ICM heating from SNe are also discussed.
This paper constitutes a generalization of a previous work (Valdarnini 2002,
 hereafter V02), 
where the investigation was mainly concerned with the numerical 
stability of different SF models. 
Section 2 presents the hydrodynamical simulations of galaxy clusters 
that have been performed. Section 3 describes the star formation 
algorithm and the modeling of ICM metal enrichment 
implemented in the simulations. Results are presented in Section 4 and
the conclusions in Section 5.

\section{SIMULATIONS}
 Here I give a short description of the simulations. Further details can be
found in V02. 

The cosmological model considered is a flat CDM model, with a vacuum energy 
density
$\oml=0.7$, matter density parameter $\omm=0.3$ and Hubble constant $h=0.7$
in units of $100 Km sec^{-1} Mpc^{-1}$. The primeval spectral index of the 
power spectrum $n$ is set to $1$ and $\Omega_b=0.015h^{-2}$ is the value of 
the baryonic density. The power spectrum of the density fluctuations has been 
normalized in order to match at the present epoch the measured cluster
number density \cite{eke96,gir98}.
Initial conditions for the cluster simulations are constructed as follows.
A collisionless cosmological N-body simulation is first run in a 
$L=200 h^{-1} Mpc$ comoving box using a P3M code with $84^3$ particles,
starting from an initial redshift $z_{in}$. At $z=0$ cluster of galaxies 
are located using a friend-of-friend algorithm, so to detect densities
$ \simeq 200 \omm ^{-0.6}$ times the background density within a radius 
$r_{200}$. The corresponding mass $M_{200}$ contained within 
this radius is defined as 
 $M_{200}= (4 \pi/3) \Omega_m \rho_c \Delta_c r_{200}^3$,
where $\Delta_c =187 \Omega_m^{-0.55}$ for a flat cosmology and $\rho_c$ is the 
critical density.
The 40 most massive 
clusters within the simulation box are identified according to this procedure 
and  
the most massive and least massive cluster (labels $00$ and $39$, respectively)
of this sample are selected for the hydrodynamical simulations.
This procedure has been already followed in V02 and the initial conditions of 
the cosmological simulation are the same. 
In addition to these two clusters, two other clusters are
considered. The original sample is enlarged to include 80 more clusters of 
decreasing virial mass. 
This amounts to a total of 120 clusters. The additional clusters extracted from 
the new sample are :
the cluster highest in mass after L$39$ (L$40$) and the least massive of the 
sample
(L$119$). Table 1 lists the properties of the four selected clusters.

\begin{table}
\centering
\begin{minipage}{140mm}
\begin{tabular}{@{}cccccc@{}}
\hline \hline
{cluster}&  {$M_{200}$} & {$r_{200}$} & {$\sigma_1$}  & $T_m$ & $f_g(0.5) 
\cdot h^{3/2}$\\
\hline
 $\Lambda$CDM00 &   $9.8\cdot 10^{14}$ & 2.01 & 1200  & 5.6 & 0.034\\
 $\Lambda$CDM39 &   $3.8\cdot 10^{14}$ & 1.5 & 800  & 3.1& 0.033\\
 $\Lambda$CDM40 &   $2.6\cdot 10^{14}$ & 1.25 & 720  & 2.8& 0.029\\
 $\Lambda$CDM119 &   $0.65\cdot 10^{14}$ & 0.8 & 480  & 1.2& 0.026\\
\hline
\end{tabular}

\caption{Reference values at $z=0$ for the four simulated
clusters used in the numerical tests. 
$M_{200}$: cluster mass within 
$r_{200}$ in units of $h^{-1} M_{\odot}$, $r_{200}$ is in units of $h^{-1}$ Mpc,
$\sigma_1$ is the central 1-D dark matter velocity dispersion in 
$Km sec ^{-1}$, $T_m$ is the mass-weighted temperature in $keV$, $f_g$ 
is the ratio of the mass of gas within the radius $r=0.5h^{-1}Mpc$ 
to the total cluster mass within that radius.}
\label{ta}

\end{minipage}
\end{table}

For each of the four test clusters, hydrodynamical TREESPH simulations 
are performed in physical coordinates. 
The initial conditions of the hydro simulations are determined 
as follows. The cluster particles at $z=0$ within $r_{200}$ of the test cluster
are located in the original simulation box back at $z_{in}$. A cube of size 
$L_c \simeq 15-25 ~h^{-1} Mpc \propto M_{200}^{1/3}$ enclosing these particles  is
 placed at the cluster centre.
A lattice of grid points is set in the cube. At each grid point is 
 associated a dark matter particle and a gas particle, of 
corresponding mass and 
coordinates. The particle positions are then perturbed, using the same initial
conditions of the cosmological simulations. 
Finally, the particles for which the perturbed positions lie inside a sphere of radius
$L_c/2$ from the cube centre are kept for the hydro simulations.
 
To model the effects of the external gravitational fields, 
the inner sphere is surrounded  out to a radius $L_c$ by
 dark matter particles with a mass eight times larger
than the sum of the masses of a gas particle and a dark matter particle of the inner sphere.
For each particle, gravitational softening parameters
are set according to the scaling $\varepsilon_i \propto m_i^{1/3}$.
The  numerical parameters of the hydrodynamical simulations are given in 
Table 2.

The simulations with index L$00$ and L$39$ have a number of gas particles 
$N_g \simeq 22,600$. For this mass resolution, the corresponding runs in V02
 have been found to give fairly stable final X-ray luminosities
for the simulated clusters. 
Additional runs have been considered, with an increased resolution with
respect to the standard runs. The high resolution (H) runs have $N_g\simeq
70,000$ and the very-high resolution (VH) runs have $N_g \simeq 210,000$.
The other numerical parameters have their values scaled accordingly.

\begin{table}
\centering
\begin{minipage}{140mm}
\begin{tabular}{@{}cccccccc@{}}
\hline \hline
{run} & {$\varepsilon_g^{~a}$} &
{$m_g^{~b}$} & {$m_d^{~c}$} & {$N_g^{~d}$} & 
{$N_d^{~e}$} &{ $N_T^{~f}$} & {$z_{in}^{~g}$}  \\

\hline
L00& 21  &$7.47\cdot10^{9}$ & $6.57\cdot10^{10}$ & 22575&25391&67388& 19.\\ 
L00H&10.5&$2.45\cdot10^9$   & $2.1\cdot10^{10}$&69599&74983&204799&29.\\
L00VH&9.9&$7.7\cdot10^8$    & $7\cdot10^9$ &212035&223235&619160&39.\\
L39 &14  &$3.7\cdot10^{9}$ & $3.22\cdot10^{10}$ &  22575 &25439&67430&19.\\
L39H &10.5  &$7.7\cdot10^{8}$ & $6.8\cdot10^{9}$ &  69599 &76255&205912&29.\\
L40 &14  &$2.3\cdot10^{9}$ & $2.1\cdot10^{10}$ &  22575 &25639&67605&19.\\
L40H &10.5  &$7.7\cdot10^{8}$ & $6.8\cdot10^{9}$ & 69599  &75447&205205&29.\\
L40VH &5  &$2.5\cdot10^{8}$ & $2.2\cdot10^{9}$ & 211954 & 224298&619918&39.\\
L119 & 10.5 &$2.2\cdot10^{9}$ & $1.5\cdot10^{10}$ &  22575 &23903&66086&29.\\
L119H & 10.5 &$7\cdot10^{8}$ & $4.9\cdot10^{9}$ &69599  &73383&203399&29.\\
\hline
\end{tabular}

\caption{Numerical parameters of the  simulations.
$^{a}$: gravitational softening parameter for the 
gas in $h^{-1}$~Kpc.  $^{b}$: mass of the gas particles in $h^{-1} M_{\odot}$. 
$^c$ : mass of the dark particles. $^d$: number of gas particles inside
the $L_c/2$ sphere at $z=z_{in}$. $^e$ : as in the previous column but for 
dark particles. $^f$: total number of simulation particles, including 
those in the external shell of radius $L_c$. 
$^g$ : initial redshift for the simulation If $z_{in} \ge 20$ the softenings are
held fixed in comoving coordinates until $z=20$, after which are kept
fixed in physical coordinates.} 
\label{tb}
\end{minipage}
\end{table}

The gravitational forces of the hydrodynamical simulations are computed 
using a hierarchical tree method with a tolerance parameter $\theta=1$ and
taking into account quadrupole corrections.
The hydrodynamical variables of the gas are followed in time according 
to the SPH Lagrangian method (Hernquist \& Katz 1989 , and references cited 
therein). In SPH, local fluid variables are estimated from the particle
distribution by smoothing over a number of neighbors.
A common choice is the $B_2$-spline smoothing kernel $W_s(r,h)$ \cite{mon85},
which has compact support and is zero for interparticle distances $|r| \ge2h$.
The smoothing kernel is normalized according to $\int W(\vec r , h) d \vec r=1$.
The smoothing length $h$ fixes the spatial resolution of the simulation.
The resolution is greatly increased in high density regions when individual
smoothing lengths $h_i$ are allowed, so that the number of neighbors of
a gas particle is nearly constant ($\simeq 32$).
 A lower limit to the smoothing lengths $h_i$ is set by $h_i \ge h_{min} \equiv 
\epg/4 $, 
where $\epg$ is the gravitational softening parameter of the gas particles.
The time integration is done allowing each particle its own time step.
The accuracy of the time integration is controlled by a number of constraints
that the individual time steps must satisfy 
( see, e.g., Valdarnini, Ghizzardi \&  Bonometto 1999).
The minimum allowed time step for the gas particles is $6.9 \cdot 10^5 yr$.

The thermal energy equation for the gas particles includes a term which models
the radiative processes of an optically thin plasma in ionization
equilibrium.
The total cooling function $\Lambda_c$ depends on the gas temperature and 
metallicity. The cooling function   
takes into account contributions from
 recombination and collisional excitation, bremsstrahlung and inverse
Compton cooling. 
Heating from an ionizing UV background has not been considered.
The cooling rate of the gas in the simulations is then 
dependent on the gas metallicity.
This is an important difference with respect the previous 
simulations (V02) and is essential in order to analyze 
 low-temperature ($T_X \simeq 2 keV$) clusters consistently.
 The dependence on the metallicity has
  indeed larger effects, as it increases the cooling rate. 
  The small scales resolved by the simulations will cool
   faster than larger ones. The 
 inclusion of cooling with its metallicity dependence 
 has then a strong impact also on the formation of larger clusters
  because of the hierarchical growth of structure.

Tables of the cooling rates as a function of the temperature and gas
metallicities have been constructed from Sutherland \& Dopita (1993)
and stored in a file. During the simulations a cubic spline interpolation
is then used to calculate 
from the tabulated values the cooling function
$\Lambda_c(T,Z)$ 
of a gas particle of given temperature $T$ and metallicity $Z$. 
Here $Z$ is the mass fraction of metals of the gas particle. 
Conversion from the metallicity $Z$ to the corresponding value of the 
iron-to hydrogen ratio $[Fe/H]$ ($[X]\equiv log_{10}X-log_{10}X_{\odot}$)
is done as in Sutherland \& Dopita.
A good fit to this relation is given by Eq. 1 of
Tantalo, Chiosi \& Bressan (1998).
The X-ray luminosities are computed from the gas emissivities within the cluster
virial radius according to the standard SPH estimator (see Eq. 8 of V02). 
The X-ray emissivity $\varepsilon(r)$
associated with a gas particle is calculated with a Raymond-Smith code
(1977) as a function of the gas temperature and metallicity.

\section{STAR FORMATION AND ICM ENRICHMENT}

Cold gas in high density regions will be thermally unstable and subject to SF.
In SPH simulations SF processes have been implemented using a variety of
 algorithms. Here conversion of cold gas particles into stars is performed
according to Navarro \& White (1993). In V02 it has been found that for 
this SF method final profiles of the simulated clusters are robust against 
the numerical resolution of the simulation.
According to Navarro \& White any gas particle in a convergent flow and for 
which the 
gas density exceed a threshold,

\be
\left\{ \brr {ll}
\nabla \cdot \vec v_i <&0  \\ 
\rho_i ~~>&\rho_{c,g}=7 \cdot 10^{-26} gr cm ^{-3},
\err
\right .
\label{eq:sfd}
\ee
will be in a collapsing region with its cooling time smaller
 than the dynamical time and
is eligible to form a star particle. If these conditions are satisfied, SF will
occur with a characteristic dynamical time scale 
$\tau_d \equiv \sqrt{3 \pi/16G\rho_i}$. The probability that a gas
particle will form a star in a time step $\Delta t$ is then given by

 \be 
 p=1-\exp {(-\Delta t /\tau_d)}.
\label{eq:ps}
\ee
 A Monte Carlo method is used at each time step to identify those gas 
particles which form star particles.
For these gas particles, a star particle is created 
with half-mass, position, velocity and metallicity of the parent gas particle.
 The star particle is decoupled from the parent gas particle
when it is created and is treated as a collisionless particle.
 A gas particle is converted entirely into a star particle when its mass 
falls below $5\%$ of the original value.
If the cooling function $\Lambda_c$ depends also on the gas metallicity $Z$, the
 density threshold criterion must be modified  in order to take into
account the increased cooling rate when $Z>0$.  
 The condition that in Navarro \& White (1993) defines
the gas density threshold $\rho_{c,g}$  now reads:
\be
\tau_{cool}= \frac {3 (\rho_{c,g}/\mu m_p) k_B T}{\Lambda_c(T,Z)}=\tau_d,
\label{tz}
\ee
where $\tau_{cool}$ is the cooling time, $T\simeq10^{6.2} \gr$,
$\mu=0.6$, $k_B$ is the Boltzmann constant and $m_p$ is the proton mass. Because of the metallicity 
dependence of the cooling function Eq. (\ref{tz}) must be solved 
numerically.
A plot of $\rho_{c,g}$ as a function of $Z$ shows that 
 $\rho_{c,g}(Z)$  is nearly flat for $Z \simlt 10^{-3}$ and has a fast decay
above $Z\simeq 5 \cdot 10^{-3}$.

It has been found that a good fit  to $\rho_{c,g}(Z)$ is  the following
 approximation:

\be
\rho_{c,g}(Z) \simeq 7 \cdot 10^{-26} gr cm ^{-3}/
(1+0.3(Z/Z_{-1.5}))^{3/2} ,
\label{rz}
\ee

where 
$Z_{-1.5}\simeq1.28 \cdot 10^{-3}$ is the metallicity corresponding to 
$[Fe/H]=-1.5$.

Once a star particle $i$ is created at the time $t_s(i)$ it will release energy into the
 surrounding gas through SN explosions. 
SN of type II (SNII) originate from the explosions of stars of mass
$m_u \geq m \geq 8 \msu$ at the end of their lifetime $\tau(m)$, 
here $\tau(m)$ is defined as in Navarro and White (1993). 

Each SN explosion produces $\varepsilon_{SNII}\equiv\varepsilon_{SN} \simeq 10^{51} $ erg ,  
 which is added to the thermal energy of the gas, and leaves 
a $\simeq1.4 \msu$ remnant.
The number of SNII explosions associated with the star particle $i$ 
in the time interval $[t-\Delta t,t]$
 is determined as 

\be
\Delta N_{II}(t)= m_s(i) \int _{\tau^{-1}(t^{\star})} ^{\tau^{-1}(t^{\star} -
 \Delta t)} 
\phi (m) dm  , 
\label{snii}
\ee
where $\phi(m)$ is the IMF of the stellar population,
$\tau^{-1}$ is the root of $\tau(m)$, $t^{\star}=t-t_s(i)=\tau$, $m_s(i)$ 
is the mass of the star particle and $m_u$ is the upper limit of the IMF 
\footnote{Hereafter masses are in solar units}. 
The normalization of the IMF is set to 
$\int _{0.1}^{m_u} \phi m dm=1$, with this normalization $N_{pop}=m_s(i)$ 
is the number of stellar populations of the star particle $i$.
Several forms of the IMF have been considered.
A Miller-Scalo (1979) has been chosen for a consistent comparison 
with the simulations of V02. For this IMF $m_u=100$. A standard IMF is
the one of Salpeter (1955), where $\phi (m) \propto m^{-(1+x)}$, with $x=1.35$.
Finally, a less steep IMF is given by Arimoto \& Yoshii (1987) for 
elliptical galaxies, for which $x \simeq 1$. The latter two IMF have $m_u=40$.
The index of the simulations with different IMF are presented in Table 3.

\begin{table}
\centering
\begin{minipage}{140mm}
\begin{tabular}{@{}cccc@{}}
\hline \hline
{run} & IMF$^{~a}$ & $\Lambda(Z)^{~b}$ & {SNIa$^{~c}$}  \\
\hline
NZIa1 & MS& No & No \\
NZIa2 & S& No& No \\
NZIa3 & A& No & No \\
NIa4 & A & Yes & No \\
S  & S & Yes & $0.07$ \\
A  & A & Yes & $0.07$ \\
\hline
\end{tabular}
\caption{Model parameters of the  simulations.
$^{a}$: shape of the IMF;  MS=Miller-Scalo,S=Salpeter and A=Arimoto-Yoshi.
$^{b}$: the cooling function includes also the dependence on the gas 
metallicity. $^c$ : SN of type Ia are considered as additional sources of the 
ICM
enrichment, a non zero entry is the value of the normalization constant
(Eq. \ref{nia}).
For these simulation $W_Z=B_2$ and $\eps=1$ (see Table 4).}
\label{tc}
\end{minipage}
\end{table}

The energy  produced in the time interval $[t-\Delta t,t]$ by the SN 
explosions of type II
associated with the star particle $i$ is $\eps \Delta N_{II}$. 
This feedback energy is returned entirely to the 
nearest neighbor  gas particles  of the star particle $i$. 
The velocity field of the neighboring gas particles is left unperturbed 
by the SN explosion, since the typical SPH simulation resolutions are
much larger than the size of the shell expansion \cite{ca98}.
The energy is smoothed  among the internal energies of the gas particles 
 according to the SPH smoothing prescription. The internal energy increment 
$\Delta u_j$ of the gas particle $j$ is then 

\be
\Delta u_j = \left ( \frac {\eps \Delta N_{II}} {m_j}   \right )
 W_s(r_j-r_i^s,h_i^s) \frac {m_j}{\rho_j} \frac{1} {V_c},
\label{wse}
\ee
where $m_j$ is the mass of the gas particle, $\rho_j$ is the gas density, 
$W_s$ is the $B_2$ SPH smoothing kernel, $2 h_i^s$ is the radius of a sphere surrounding
$N_s \simeq 32 $ gas neighbors of the star particle $i$ and 
$V_c = \sum_j W_s(r_j-r_i^s,h_i^s) m_j/\rho_j $ is a normalizing  factor.
$V_c$ has been introduced to avoid  that 
$\sum_j \Delta u_j m_j \ne \eps \Delta N_{II}$.
The smoothing length $h_i^s$ is constrained by the upper limit 
$h_i^s \le 30 Kpc=h_M$ 
and there is no lower limit, unless explicitly stated.  
 In high density regions the SN energy is then returned to the gas neighbors 
without being affected by the SPH resolution length.
The choice of the value of $h_M$ is discussed
in sect. 4.2, where the dependence of the 
simulated cluster profiles on a number of parameters is analyzed.

The number of SN of type Ia (SNIa) associated with the star particle $i$ in
the time interval $[t-\Delta t,t]$  has been determined according to
Greggio \& Renzini (1983). The scheme implemented follows 
Lia, Portinari \& Carraro (2002). The number of SNIa events at the epoch $t$ 
 is

\be
N_{SN Ia}(<t) = m_s(i)~A \int_{m_{b,inf}}^{m_{b,sup}} \phi(m_b) 
\left[ \int_{\mu_{inf}}^{0.5} f(\mu) d\mu \right] dm_b,
\label{nia}
\ee
here $m_b$ is the mass of the binary system,  
 $0 \le \mu=m_2/m_b \le 0.5$ is
the mass fraction of the secondary and $f(\mu)=24 \mu^2$.
The mass $m_b$ of the binary system lies in the range 
 $3 =m_{b,inf} \le m_b \le m_{b,sup}= 12 =2M_{up}$, $m_2=m_2(\tau)$ is the 
mass of the secondary which ends its lifetime at the time $t=\tau(m_2)+t_s(i)$ 
and
$\mu_{inf}=MAX\{ m_2(\tau)/m_b,1 -M_{up}/m_b\}$. The lower limit 
to the mass of the secondary, $0.9 \le m_2$, is given by the age of the 
universe.
The normalization constant 
$A$ is set to $A=0.07$ in order to match the estimated rate of type Ia
SNe in galaxies \cite{pcb98}. Inverting the order of integrations one
obtains:

\be
N_{SN Ia}(<t) = m_s(i) ~A  \int_{m_2(\tau)}^{M_{up}} 
		 24 m_2^2 \left [ \int_{m_{b,min}}^{m_{b,max}}  
		 \frac{\phi(m_b)}{m_b^3} dm_b \right ] dm_2 ,
\label{snia}
\ee
where $m_{b,min}=MAX \{ m_{b,inf}, 2 m_2 \}$,
$ m_{b,max}=MIN\{ m_{b,sup}, m_2+M_{up}\}$ for the number of SNIa 
 explosions associated with the star particle $i$ at the age $t$.
The number of SNIa explosions 
 in the time interval $[t-\Delta t,t]$  is then 
$ \Delta N_{Ia}=\left ( N_{SN Ia}(<t)-N_{SN Ia}(<t-\Delta t) \right )$.
The corresponding explosion energy is $\varepsilon_{SNIa} \Delta N_{Ia}$,
with $\varepsilon_{SNIa} \simeq \eps$ \cite{woo86}.
This energy is smoothed as in Eq. (\ref{wse})
over the nearest neighbor gas particles of the star 
particle $i$.

SN explosions also inject enriched material into the ICM,
thus increasing its metallicity with time. The mass of the $k-th$ heavy 
element produced in a SN explosion is defined as the stellar yield
$y_{S,k}(m)$, with $S$=II or Ia. The total yield is the sum over the 
masses of the heavy elements : $y_S(m)=\sum_k y_{S,k}(m)$.
For SNIa the yield $y_{Ia,k}$ is a constant which is independent on the
progenitor mass.
The adopted SNIa yields are those of Iwamoto et al. (1999, W7 model),
with $y_{Ia}=1.4 $ and $y_{Ia,Fe}=0.7$.
The yields $y_{II,k}$ of type II SN depend on the progenitor mass $m$ 
and it is useful to define an average yield

\be
<y_{II,k}>=  
\frac{\int_{m_l}^{m_{u}} y_{II,k}(m)\phi(m) dm } 
{\int_{m_l} ^{m_{u}} \phi(m) dm} .
\label{yii}
\ee
The upper limit $m_u$ is defined according to the IMF and $m_l$ is the 
lower bound for SNII progenitors. Here it is assumed $m_l=8$.
For type II  SNe the predicted theoretical yields suffer from a number of
uncertainties. According to the assumed massive star physics the yields of
different models can differ by a factor $\simeq 2$ or more.
A detailed discussion can be found in Gibson, Loewenstein \& Mushotzky (1997).
The theoretical model chosen here is model B of Woosley \& Weaver (1995).
For this model the explosion energy of massive stars is boosted and 
the produced metals are not reabsorbed in the SN explosion.
The yields of this model are therefore the most favorable from the 
point of view of the amount of iron synthesized from SNII.
Final metal abundances for a different theoretical SN explosion model
 can be obtained from the SNII component of the gas metallicity profile 
by a rescaling of the yields. 
This is a valid approximation as long as there are small differences 
between the yields $y_{II}$ which determine the gas metallicity $Z$, 
and therefore the local cooling rate and the SF threshold (Eq. \ref{rz}). 

Woosley \& Weaver (1995) yields are available down to $m_l=11$.
The yields between $m_l=8$ and $m=11$ have been obtained with a 
linear extrapolation, the associated errors are however small
since SN nucleosynthesis is negligible  below $m=11$.
From Table 1 of Gibson, Loewenstein \& Mushotzky (1997) 
$<y_{II,Fe}>\simeq 0.14$ for the 
Woosley \& Weaver B model with a Salpeter IMF, $m_l=10$ and $m_u=50$.
 I obtain the same value for an Arimoto-Yoshi IMF with $m_l=8$ and 
$m_u=40$.
The mass of heavy elements which has been injected into the ICM 
at the age $t$ due to the SNII explosions associated with the star
particle $i$ is determined as \cite{ti80} :

\begin{table}
\centering
\begin{minipage}{140mm}
\begin{tabular}{@{}ccccccc@{}}
\hline \hline
{run} & IMF & SNIa & $h_M^{~a}$ & $W_Z^{~b}$ & $\eps^{~c}$ & $h^{d}_{min}$ \\
\hline
A & A& 0.07 & 30 & $B_2$ &1 &0\\
A2 & A& 0.07 & 10 & $const$ &1 &0\\
A3 & A& 0.07 & 30 & $const$ &1 &0\\
A4 & A& 0.07 & 30 & $const$ &1/10&0 \\
A5 & A& 0.07 & 30 & $const$ &1&12 \\
A6 & A& 0.07 & 30 & $const$ &1& 6\\
\hline
\end{tabular}
\caption{ Additional model parameters for runs with AY IMF (label A).
$^{a}$: maximum smoothing length for the metal enrichment
spline.  $^{b}$: Functional form of the chosen spline for the metal enrichment
of the gas;
$B_2$: standard SPH, $const$ for a uniform spread. 
$^c$ : SN explosion energy in units of $10^{51} erg$.
$^d$ : minimum smoothing length for the metal enrichment.}
\label{td}
\end{minipage}
\end{table}

\be
m_{s,ZII}(<t)= m_s(i) \left \{ 
\int_{m(\tau)}^{m_{u}} \left[ m-m_r-y_{II}(m)\right] Z_s(i) \phi(m) dm + 
\int_{m(\tau)} ^{m_{u}} y_{II}(m) \phi(m) dm 
\right \},
\label{mzej}
\ee
where $Z_s(i)$ is the metallicity of the star particle $i$ ,
$m(\tau)$ is the SNII progenitor mass at the epoch $\tau=t-t_s(i)$ and
$m_r\simeq1.4$ the remnant mass. Similarly, the total mass returned to the ICM
is 

\be
m_{s,ejec}(<t)= m_s(i)  \int_{m(\tau)}^{m_{u}} \left[ m-m_r \right] \phi(m) dm.  
\label{mej}
\ee
The ejected masses from star particle $i$  in the time interval 
$[t-\Delta t,t]$ are then $\Delta m_{s,esp}=m_{s,esp}(t)-m_{s,esp}(t-\Delta t)$,
with $esp= ZII$ or $ejec$.
For each star particle $i$, these masses are calculated at each time step
according to the stellar particle age $\tau$. The $\Delta m_{s,esp}$  are then 
distributed over the gas neighbors of the star particles, using the same
smoothing procedure adopted (Eq. \ref{wse}) for returning the SN feedback energy
to the gas, i.e. 

\be
\Delta m_{ZII}(j) =   \Delta m_{s,ZII}  
 W_Z(r_j-r_i^s,h_i^s) \frac {m_j}{\rho_j} \frac{1} {V_c},
\label{wze}
\ee
where $\Delta m_{ZII}(j)$ is the mass increment of the metallicity of type II 
of the gas particle $j$, neighbor of the star particle $i$, and
$W_Z$ is the smoothing kernel.
 The same relation holds for the total ejected mass $\Delta m_{s,ejec}$.
This smoothing procedure is similarly applied  to calculate the metallicity
increment of the gas particles due to SNIa ejecta.  
Since $y_{Ia}$ is a constant for SNIa the mass increment, $\Delta m_{s,ejec}$
is defined as $\Delta m_{s,ejec}= \Delta N_{Ia}(t)~ y_{Ia}$.
Each gas particle, as well as a star particle, has two different metallicity 
variables. These variables represent, respectively,
the total mass of metals originated from SNII and SNIa events.
This is necessary in order to discriminate 
the different contributions of SNII ejecta from that of SNIa
in the final profiles of the gas metallicity. 
The above integrals are calculated numerically as a function of the progenitor
mass $m=\tau^{-1}(t)$ and stored in tables.  A linear interpolation procedure
is used during the simulation to obtain from the tables the values of the 
integrals at the age $t$.

Since the choice of the smoothing kernel $W_Z$ is somewhat arbitrary, 
a standard choice is the SPH smoothing kernel  \cite{mo01,li02}. 
However, if the ejected material is deposited at the end of the shock expansion,
a less steep deposition profile is a better description of the deposition
mechanism. For a uniform distribution, $W_Z \propto const$.  
Aguirre et al. (2001) have investigated the metal enrichment of the diffuse
intergalactic medium in cosmological simulations. They have considered 
a power-law shape $W(r,h) \propto r^{\alpha}$ 
for the deposition kernel. 
They assumed a 
default value $\alpha=3$, the uniform case considered here corresponds 
to their $\alpha=2$.
The simulations have been performed using for $W_Z$ the SPH smoothing kernel
as the default kernel.
 Table 4 reports the index of the runs for which $W_Z=const$.
Each simulation has a label obtained by merging together the labels
of Table 2 and 3, or Table 2 and 4, which determine the simulation parameters.

\begin{figure}
\vfill
\centerline{\mbox{\epsfysize=14.0truecm\epsfxsize=14.0truecm
\epsffile{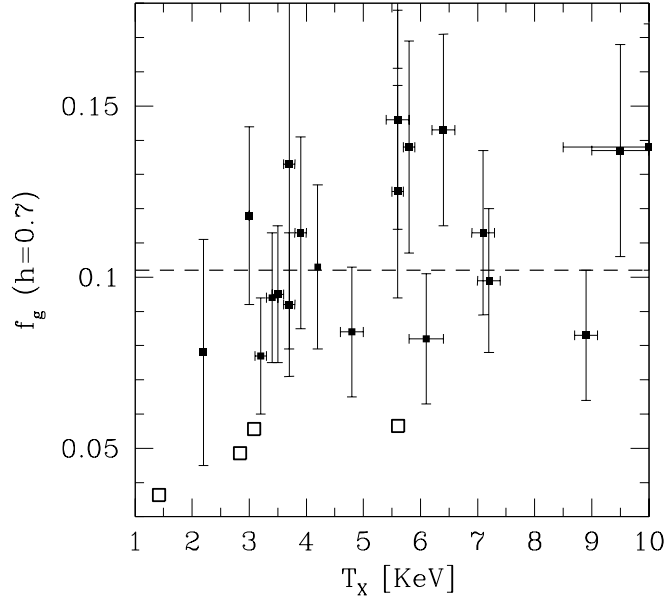}}}
\caption{ 
Baryonic fraction for the Mastumoto et al. (2000) sample of nearby ($z<0.1$)
clusters. $f_g=M_g/M_T$ is evaluated at $r=0.5h^{-1}Mpc$. Open symbols  refer to
the simulated clusters. The dashed line is the cosmological value 
($\Omega_b/\Omega_m$) for the assumed model.}
\label{fg}
\end{figure}

\section{RESULTS}
A measure of the regularity of the numerical cluster sample can be assessed
from a comparison of the mean gas fraction of the 
four test clusters $f_g=M_g/M_T$ within the radius $r=0.5h^{-1} Mpc$,
with the corresponding values of the Matsumoto et al. (2000) sample
(Fig. \ref{fg}). This sample has been chosen for a comparison of the 
global cluster iron 
abundances with the ones predicted by several runs (see below).
The sample values of $f_g$ shown in Fig. \ref{fg} have a large scatter 
around the cosmological value of the model $f_b=\Omega_b/\Omega_m=0.1$, while
those of the simulated clusters are systematically smaller.
This indicates that the simulated clusters are regular objects 
\cite{ev96} in a quiet dynamical state, without having undergone 
recent mergers, a conclusion supported also from previous substructure 
analyses of 
X-ray maps (Valdarnini, Ghizzardi \&  Bonometto 1999) for a 
$\Lambda$CDM cluster sample with the same cosmological initial conditions.
Therefore, the profiles of simulations with different model parameters 
can be consistently compared without biases that can follow from the 
presence of substructure.
\subsection{Simulations with different IMF and cooling}

Results from simulations with different IMF and cooling parameters are
discussed first. The relevant parameters for these runs are presented 
in Table 3. The simulation results are shown here only for the cluster
$\Lambda$CDM39; for the other three test clusters 
there are not qualitatively relevant differences
in the final profiles. 
The simulations have been performed keeping the numerical
resolution, which is given by the index L39 of Table 2, fixed.
The first three runs  of Table 3 (L39NZIa1, L39NZIa2 and L39NZIa3) correspond
to the following choices of the IMF : Miller-Scalo (MS),
Salpeter (S) and Arimoto-Yoshi (AY), respectively. All the other parameters 
have been left unchanged. 
These simulations do not consider the possible dependence of the cooling 
function on the gas metallicity and the contribution of SN of type Ia to the 
gas metal enrichment.
For this choice of parameters, the run L39NZIa1 is just case cl39-10
of V02, in order to compare final profiles with previous findings.
L39Ia4 has the metallicity dependence of the cooling function switched on.
The other two indexes of Table 3 (S and A)  are for runs with SNIa as a heating
source of the ICM, as well as of metal enrichment.

\begin{figure}
\vfill
\centerline{\mbox{\epsfysize=14.0truecm\epsfxsize=14.0truecm
\epsffile{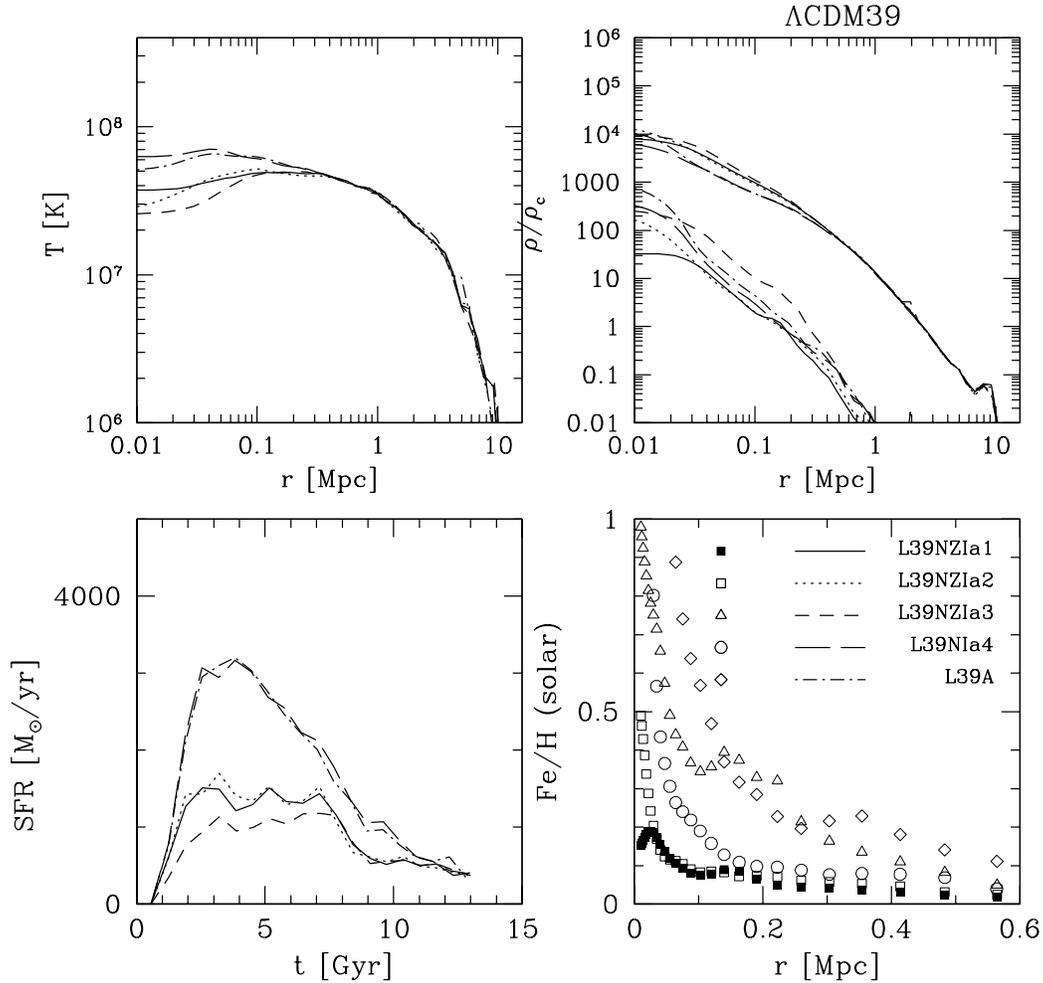}}}
\caption{ 
For the cluster $\Lambda$CDM39, a comparison of simulations with different 
choices of the cooling function and the IMF. 
Bottom left: SF rates as a function of time. The other panels refer to
$z=0$. From the top left : temperature, gas densities and iron abundances 
profiles versus radial distance. Iron abundances are in solar units.
 In the density panel the lower profiles are those of the metals.
The meaning of the simulation labels is given in Table 2 and 3.}
\label{NZ}
\end{figure}

Fig. \ref{NZ} shows the final profiles of temperature,
gas and iron abundances in solar units for the different runs. 
 The SF rate (SFR) as a function of age 
is also plotted.
While a comparison of the iron abundances obtained from simulations with 
available data will be discussed later, several conclusions about the choice 
of the IMF
can already be drawn from the iron profiles seen in Fig. \ref{NZ}.
For a cluster like \La39, the iron abundance expected at a distance 
$\simeq 0.4 Mpc $~$( \simeq 0.2 r_{200})$ from the centre is in the range 
$\simeq 0.3 -0.35$ \cite{de01}. Hereafter iron abundances  are given is solar 
units ($Fe/H=4.68 \cdot 10^{-5} $ by number).
The iron profiles of Fig. \ref{NZ} show that a Miller-Scalo IMF is completely
ruled out as a possible IMF and cannot produce the amount of iron  
required by observations. The same conclusion holds also for a Salpeter IMF,
for which $Fe/H \simlt 0.1 $ at $r\simeq 0.4 Mpc$. The only IMF for
which the simulation gives a significant amount of iron is Arimoto-Yoshi
(AY). The iron profile of the S run L39NZIa2 is always a factor 
two below that of L39NZIa3 (AY).
At $r\simeq 0.4 Mpc$ the latter profile gives $Fe/H \simeq 0.2-0.15$.
These values are still below the ones required by observations, 
nevertheless it appears that the AY IMF gives the best results from the 
point of view of the amount of iron required to fit the data.
These conclusions are in agreement with those of 
Loewenstein \& Mushotzky (1996).

These iron profiles are originated from the metal ejecta of SN II in the ICM,
with the bulk of the SN explosions that has enriched the ICM already at 
$z\simeq 0.7$. Therefore, the gradients in the final iron profiles indicate
that the cluster  was already dynamically relaxed at this epoch, without 
major mergers of substructure which could have remixed the ICM and erased
the original gradients.
The most important differences in the profiles arise when the cooling function
also depends on the gas metallicity (L39NIa4). 
For this run, the SFR is much higher than in the previous cases. For the range
of temperatures considered here, this is a direct consequence of the 
higher cooling rate. As a consequence, the density threshold criterion for
SF is lower and the SF activity is much higher than in the no-metal cases.
 The top right panel of Fig. \ref{NZ} shows that final gas density profiles are
 not strongly affected by the metallicity dependence of the cooling rate.

One of the most important consequences of considering metallicity effects for 
the cooling rate is seen in the temperature profiles.  From the top left panel 
of 
Fig. \ref{NZ} there is clear evidence that the temperature profile of 
L39NIa4  has a different shape from the other profiles already at
$r\simeq 200 Kpc$ from the cluster centre.
Between this distance and the cluster centre, the profile shows values of
the temperature higher than in the runs with the no-metal cooling.
The metallicity dependence of the cooling rate affects   
 the final temperature profiles at the cluster centre through two main effects. 
The first is the term $\Lambda_c(T,Z)$, which enters in the gas thermal
equation. Even for gas temperatures above $ \simeq 2keV$ 
the increase in the cooling rate with respect the no-metal case can
be significant for high metallicities ($Fe/H \simgt 1$), which can be
present at the cluster centre if there are strong metallicity gradients.
The second effect is indirect; because of the increased 
cooling rate, SF has been higher in the past than in the no-metal case,
and as a consequence there is a larger amount of cold gas that has been 
converted into stars and a deeper potential well at the cluster centre.
As a result, there is a higher inflow at the cluster centre of the 
surrounding high-entropy gas than in the runs that do not consider 
the metallicity dependence of the cooling function. This in turn implies 
higher temperatures at the cluster centre (see sect. 4.2).
An important result that therefore follows from these simulations is that
the temperature profiles in the cluster inner regions cannot be considered 
as approximately flat. This is in disagreement with what has been found in V02, 
where the simulations  included radiative cooling and SF, but did not take into
account the metal contribution to the cooling.
These conclusions are rather general and are valid not only for \La39, which 
has a mass-weighted temperature $T_m\simeq 3keV$, but also for a cluster like
\La00, for which $T_m \simeq 6keV$ (see Fig. \ref{lx} below).

The iron abundance profile of the run L39NIa4 is lower by a 
factor $\simeq 50 \%$ with respect that of L39NZIa3, where the cooling
function does not depend on the gas metallicity. 
This is also confirmed by contrasting the metal density profiles in the top
panel of Fig. \ref{NZ}.
 If the cooling rate increases when the metallicity dependence is taken 
into account, then the final amounts of stars and metals at the cluster centre
are expected to be higher than in the runs with the no-metal cooling.
This is in contrast with what is found.
It is not clear how this anti-correlation 
 between the final metal ICM abundance and the metal dependence of the 
cooling rate originates.
A possible explanation lies in the strong SF activity that occurs earlier 
for the run under consideration.
 As a consequence, most of the ICM is metal enriched by
stars at earlier epochs.
An iron profile that is in better agreement with data is recovered when 
SNe of type Ia (L39A) are also considered as sources of metal enrichment 
for the ICM.
For this run, the iron abundance at $\simeq 300 Kpc$ is $\simeq 0.2$ , 
about two times that of the parent run with the same parameters but without 
SNIa (L39NIa4). This is obtained for $A=0.07$ (Eq. \ref{nia}). According to
Portinari, Chiosi \& Bressan (1998) this choice of $A$ corresponds to a 
number of SNIa $\simeq 20 \%$ of the total number of SNe in the Galaxy.
A numerical simulation with $A=0.04$ has shown that the amount of iron that is 
injected from
SNIa  in the ICM is roughly proportional to $A$.
The above results indicate that for an AY IMF 
as much as $\simeq 50\%$ of the iron
in the ICM comes from SNIa, provided that the current rate is normalized to 
match that estimated from spiral galaxies.

\begin{figure}
\vfill
\centerline{\mbox{\epsfysize=8.0truecm\epsfxsize=14.0truecm
\epsffile{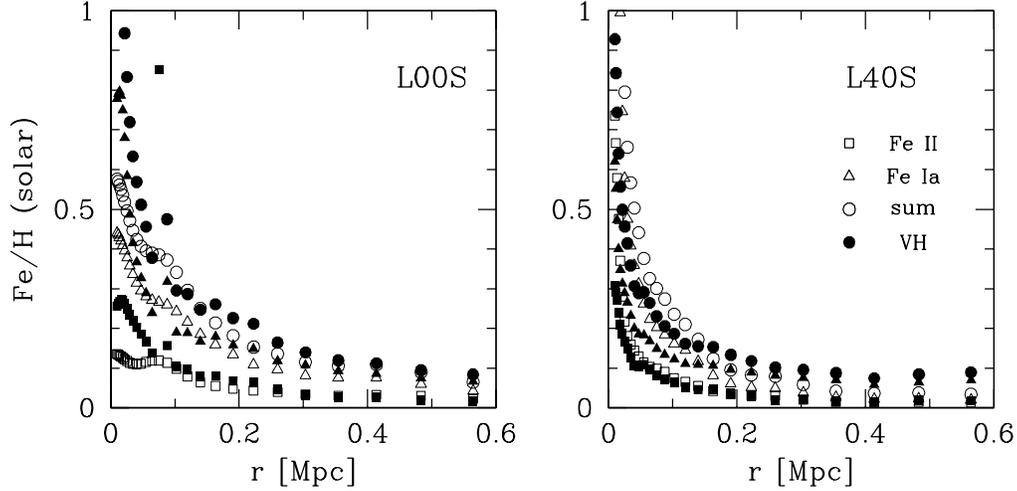}}}
\caption{ 
For the clusters $\Lambda$CDM00 and $\Lambda$CDM40 final iron profiles from 
SN II/Ia 
are displayed for simulations with a Salpeter IMF (see Table 3).
Filled symbols are for the corresponding very-high resolution runs.}
\label{VHS}
\end{figure}

The stability of the final iron profiles against the simulation's  numerical
resolution can be estimated from Fig. \ref{VHS}, where for a chosen IMF (S) the
iron profiles of the two clusters \La00 and \La40 are displayed for the 
standard resolution and the very-high resolution runs (VH).
There is a good agreement between the profiles with a different resolution,
with small differences localized within $r \simlt 0.1Mpc$.

\subsection{Simulations with different metal ejection parameters}
Values of the global iron abundances $A_{Fe}=M_{Fe}/M_H$ are listed in solar
units in Table 5 for the various simulations. These values have been
calculated at a fiducial radius $r=0.5 h^{-1}Mpc$ and are in the range 
$\simeq 0.15-0.2$ for the simulations with index A. From the Matsumoto 
et al. (2000) sample of nearby clusters ($z<0.1$) the estimated abundances
at $0.5 h^{-1}Mpc$ are all above $\simeq 0.2$ and in the range $\simeq 0.2-0.4$
(see later Fig. \ref{lx}).
Therefore, it is important to investigate the effects on the final iron 
abundances of adopting prescription of metal ejection different from that 
of the A runs, keeping fixed the other parameters of the simulations.

The largest amount 
 of iron ejected in the ICM are obtained for the AY IMF, and this is the IMF 
that is chosen 
in the remaining of the paper in order to 
account for the measured iron abundances.
Hence for this IMF, different model parameters of metal ejection
have been tested, Table 4 presents the index of the runs and the
associated relevant parameters.

The metal enrichment of the gas is modeled according to Eq. \ref{wze},
with the mass that is ejected by a star particle $i$ in the interval $\Delta t$ 
distributed over the gas neighbors, the mass fraction is weighted according to 
the smoothing kernel $W_Z(r,h)$.
For the A runs the kernel $W_Z$ is the standard SPH $B_2$ spline.
The radial profile of the deposition kernel is one of the parameters that 
governs the distribution of the ejected metals among the gas particles
surrounding the star particle.
According to Aguirre et al. (2001) the form of the distribution function 
$W_Z(r,h)$ can be generically assumed with a radial power law $\propto 
r^{\alpha}$ behavior. A radial profile shallower than that of the $B_2$
spline clearly implies that more metals are deposited  near the
limiting radius. Therefore, gas particles that are not part of a SF activity
can be metal enriched and diffuse metals into the ICM. 
The metal enrichment of the ICM is expected to be
 higher in  this case because metals are more effectively mixed.
This has been also
pointed out by Mosconi et al. (2001). 
Alternatively to the $B_2$ spline, the shape of the deposition profile
chosen here is $W_Z=const$, which corresponds to the case of a uniform metal
distribution.

\begin{figure}
\vfill
\centerline{\mbox{\epsfysize=14.0truecm\epsfxsize=14.0truecm
\epsffile{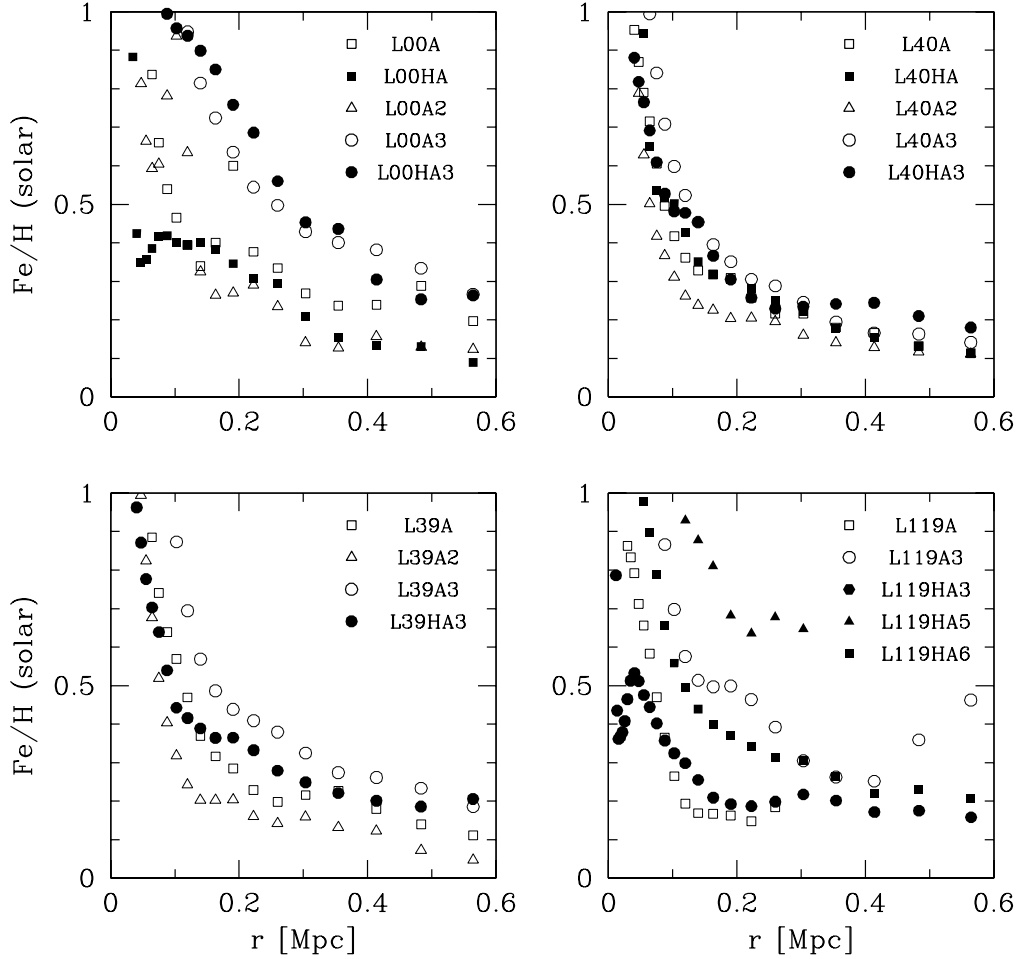}}}
\caption{ 
For the four test clusters final iron profiles are plotted for runs 
with different model parameters of metal enrichment (Table 4)
and numerical resolution.}
\label{A3}
\end{figure}

The simulations with index A3 in Table 4 have $W_Z=const$ and are the mirror
simulations of the runs with index A, the only different parameter being the 
choice of the deposition kernel.
Final iron profiles are compared against those of models A in Fig. \ref{A3}.
The four panels of Fig. \ref{A3} refer to the four test clusters.
In each panel final iron profiles are plotted for the runs with 
different metal ejection parameters and different numerical resolutions.
The plots of Fig. \ref{A3} show that, for models A3, the profiles are higher
than that of models A by a factor that is in the range $\simeq 1-2$.
The differences are largest for \La00 and \La39, modest for \La40.
The global iron abundances $A_{Fe}$ are now in the range $\sim 0.23 - 0.35$ 
(Table 5),
in better agreement with data than the values of models A.
These results demonstrate that the choice of the deposition profile is a key
parameter in determining iron abundances in the ICM.
The differences in the final profiles between the runs A3  and A  gives a
measure of the scatter in the final abundances that is inherent to the choice
of different kernels.

It must be stressed that other choices are clearly possible; in particular
a steeper IMF may even require a deposition kernel with a positive radial 
derivative.  A quantitative comparison with several data of the profiles
obtained from model A3 for the four test clusters is  performed
later. The results show that for models A3 the simulations  
are in good agreement with the measured values and the parameters of the A3 
runs are 
then taken here as those of a `fiducial model' against which to compare 
the results from the other simulations.

The stability of the simulated iron profiles  against the 
 numerical resolution is tested in Fig. \ref{A3} by plotting (filled symbols)
profiles of high-resolution simulations (H in Table 2) against the corresponding
ones with standard resolution.
As can be seen, there are no large differences in the final profiles
between the high and standard resolution runs. The only important 
exception is for the coldest cluster \La119 ($T_m \simeq 1.5 keV$). 
The iron profile of L119HA3
is much lower than that of the standard resolution run L119A3.
This strongly suggests that there are numerical resolution problems
that affect the profile of \La119, when the numerical resolution is that 
of the standard runs.  Increasing the numerical resolution implies that 
smoothing lengths are smaller. As a consequence the metals are distributed 
among gas neighbors in smaller volumes, this implies a smaller mixing of 
metals in the ICM (see also Mosconi et al. 2001).
Clearly, this effect should not be seen (cf. Fig. \ref{VHS}) 
and if present implies that low-resolution runs are undersampling the gas 
distribution.
In order to reliably predict metallicity profiles it is then safe to assume
that simulations of low-temperature ($T\simlt 3 keV$) clusters require at 
least the mass resolution of high-resolution runs.
To explain the low values of the $A_{Fe}$ for the run L119HA3 an alternative 
hypothesis is that the metal enrichment of the ICM by SNe is characterized
by a minimum diffusion length.
 Simulations have been performed without a lower limit $h_{min}^s$ for the
smoothing lengths $h^s_i$ of the diffusion kernel $W_Z$ of ejected 
metals.
For cool clusters high resolution runs may therefore imply values of $h^s_i$ 
in the central high density regions below $h_{min}^s$.
This in turn would imply lower values for the metallicity profiles because
of the reduced mixing of the ejected metals in the ICM.
This effect is very similar to the behavior expected when the numerical
resolution is insufficient and high-resolution runs yield lower profiles 
than those of low-resolution runs.
The profiles of Fig. \ref{VHS} show that this effect is relevant only for the 
less massive, cool, clusters ($\simlt 3keV$). For the numerical parameters 
of the high-resolution runs (H) this implies that $h_{min}^s$ must be of the 
order of $\sim $ few $Kpc$. In order to investigate the dependence of the final 
metallicity profiles on the value of $h_{min}^s$ two simulations have
been performed for the cluster \La119.
Models A5 and A6 have the same parameter of model A3, but with $h^s_{min}=12$
 and $6 Kpc$, respectively.
Iron profiles for these two runs are plotted in the panel of Fig. \ref{A3}
for the cluster \La119. The results demonstrate that requiring a minimum
diffusion length is highly effective to obtain for cool clusters a 
large-scale mixing of iron in the ICM. The iron abundances $A_{Fe}$ are now 
in the range $\sim 0.4-0.5$, in better agreement with the observational
values for cool clusters.

 Another parameter which is important to determine the amount of
metal enrichment of the ICM is the upper limit $h_M$ (see Table 4) of the 
smoothing lengths $h_i^s$. These are defined in Eq. \ref{wse} as the 
half-radius of a sphere enclosing $\simeq 32 $ gas neighbors of the generic
star particle $i$. The standard value assumed for the simulations is
$h_M=30 Kpc$ \footnote{This is the same value used in V02;
 the value quoted in the text of V02 is erroneously half the true value}.
The sensitivity of the final profiles to the assumed values can
be estimated from the plots of Fig. \ref{A3}. For the runs A2, $h_M=10Kpc$.
Contrasting the profiles with those of the A3 runs shows clearly that models A2 
yield final abundances well below those of models A3 and comparable
to the abundances of models A.
For the runs A2, the abundances $A_{Fe}$ are of the order of $\simeq 0.1-0.13$,
 a factor $\simeq$ two smaller than the values of runs A3. 
The value of $h_M=10Kpc$ is clearly ruled out by the measured iron 
abundances. The choice $h_M=30 Kpc$ gives much better results for the metal 
distribution in the ICM.
In fact, the values of $h_i^s$ are rarely fixed by this upper limit.
A radial binning of the squared distribution $(h_i^s)^2$ shows that on average 
the rms values of the smoothing lengths $h_i^s$ are below $10 Kpc$ in 
high-density regions ($r \simlt 0.1Mpc$), and grow up to $\simeq 20Kpc$ 
in low-density regions ($ r \simeq 500 Kpc$) for which $\rho/\rho_c \simeq 10$.
 This shows that the choice of the value of $h_M$ has a relevant
impact on the final metallicity profiles at radial distances $r\simgt0.1Mpc$,
in low-density regions where is most likely that gas particles that are not in
 a SF region can get metal enriched and diffuse metals.
These values of the smoothing lengths  
correspond to maximum radii $\simeq 40Kpc$ and are 
much higher than the upper limit of $\simeq 10Kpc$ estimated by 
Ezawa et al. (1997) for the diffusion of ions in a Hubble time in a 
low-density plasma ($\rho \simeq 10 \rho_c$) at a temperature of $\simeq 4keV$.
On the other hand the Ezawa et al. limit corresponds here to $h_M=5 Kpc$, as it has been found for models A2 this choice of $h_M$ would imply for the 
simulated clusters final 
iron abundances in the ICM much lower than the measured values.
The plots of Fig. \ref{VHS} \& \ref{A3} show also that for 
clusters with temperatures above $\simgt 2keV$ 
the profiles of model A3, with $h_M=30Kpc$, do not depend on 
the numerical resolution of the simulations.
A possible way of reconciling these discrepancies lies in the fact that
analytical estimates of the maximum diffusion length of metals in the 
ICM do not take into account the mixing of metals that can occur because
of the dynamical interactions between cluster galaxies and the ICM
\cite{dub20}.
Simulation profiles show that to model the mixing of metals the  best results 
are obtained for the parameters of model A3. 

\begin{table}
\centering
\begin{minipage}{140mm}
\begin{tabular}{@{}ccc@{}}
cluster&runs&${\bar A}_{Fe}(0.5)$\\
\hline \hline
$\Lambda$CDM00& L00S/L00VHS/L00A/L00HA/L00A2/L00A3/L00HA3&
0.106/0.145/0.236/0.251/0.147/0.365/0.376\\
$\Lambda$CDM39&  L39A/L39A2/L39A3/L39HA3&
0.188/0.127/0.288/0.229\\
$\Lambda$CDM40& L40S/L40VHS/L40A/L40HA/L40A2/L40A3/L40HA3&
0.08/0.106/0.208/0.203/0.185/0.277/0.245\\
$\Lambda$CDM119& L119A/L119A3/L119HA3/L119A4/L119HA4/L119HA5/l119HA6&
0.186/0.377/0.181/0.392/0.147/0.475/0.283\\
\hline
\end{tabular}
\caption{ Average value in solar units ($4.68\cdot 10^{-5}$) of the iron 
abundance ${\bar A}_{Fe}=M_{Fe}(<r)/M_{H}(<r)$ for the  considered runs. 
${\bar A}_{Fe}$ is evaluated at $r=0.5 h^{-1}Mpc$.}
\label{te}
\end{minipage}
\end{table}

\subsection{Comparison with data}
\begin{figure}
\vfill
\centerline{\mbox{\epsfysize=12.0truecm\epsffile{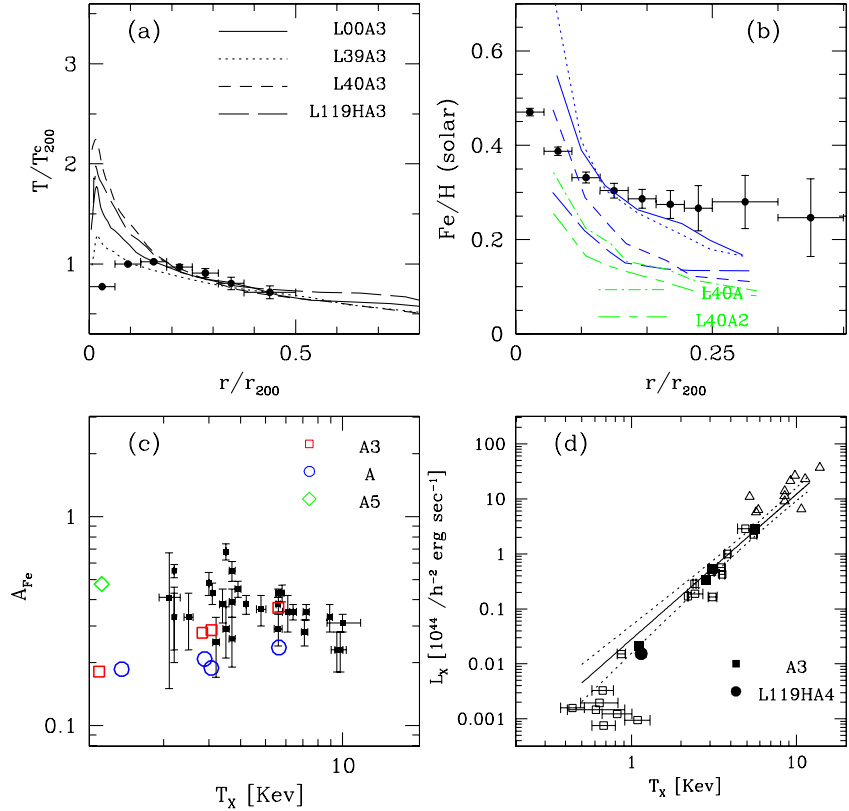}}}
\caption{ 
Comparison against data of several observational variables at $z=0$ from the
 runs with an Arimoto-Yoshi IMF (A). 
{\bf (a)}: The projected emission weighted temperature profiles as a
function of $r/r_{200}$ for the runs A3 (Table 4); data points from 
De Grandi \& Molendi (2002), $T^c_{200}$ is $T_{200}$ normalized to match
the last data point ($T_{200} \simeq 80 \% T^c_{200}$).
{\bf (b)}: As in panel (a), but for the projected metallicity profiles;
data points from the CF clusters of De Grandi \& Molendi (2001). Blue lines
are the profiles corresponding to the runs of panel (a).
{\bf (c)}: Average iron abundances at $r=0.5h^{-1}Mpc$, open squares 
refer to A3 runs and open circles to A runs; filled squares from the 
nearby cluster sample of Matsumoto et al. (2000).
{\bf (d)}: Bolometric X-ray luminosities as a function of the temperature.
Data points from Fig.11 of Tozzi \& Norman (2001). For the sake of clarity
only a fraction of the data set has been plotted. The continuous line is 
the best-fit $L_X=3.11\cdot 10^{44} h^{-2} (T/6keV)^{2.64}$ of Markevitch
 (1998),
dashed lines are the $68\%$ confidence intervals. 
Filled squares are the values of $L_X$ for the A3 runs, the filled 
circle refers to the A4 run ($\varepsilon_{SN}=10^{50} erg$).
As in Markevitch, $L_X$ for the simulated clusters has been estimated 
by removing a region of radius $50h^{-1} Kpc$ centered on the peak of the 
gas density. Mass-weighted temperatures have been used as estimators
of the spectral-fit temperatures.
}
\label{lx}
\end{figure}

For the four test clusters, observational variables from simulations with
different model parameters are compared in Fig. \ref{lx} against a number
 of data. For model A3 the projected emission  weighted temperature profile
are plotted as a function of radius in panel (a). Data points are the
mean error-weighted temperature profiles of 11 cooling flow clusters from 
De Grandi \& Molendi (2002). The profiles have been calculated according 
to Eq. A3 of De Grandi \& Molendi. 
  The cluster centre is defined as the maximum of the gas density,
 and a peak emission criterion for defining the centre does not modify the 
calculated profiles in a significant way.
Each smoothed profile has been rescaled to match the last data point.
There is a remarkable agreement of the simulated profiles with data, the 
only important exception being for the two innermost bins.
The temperature profiles show a radial increase toward the cluster central 
region, followed by a strong drop at the cluster centre.
This feature is common to all the runs and is not shared by the data, for
which the innermost bin has a lower temperature than the nearby ones.
This behavior of the temperature profile is robust and is not sensitive 
to an increase of the numerical resolution. On the other hand, this increase
of the cluster temperature at the centre is a consequence of the entropy
conservation during the galaxy formation and the subsequent removal of
low-entropy gas \cite{wu02a}.
A possible explanation for this discrepancy lies in the fact that what is
being measured are spectral temperatures.
Mathiesen \& Evrard (2001) argue that spectral fit temperatures are weighted by
the number of photons, so line cooling from small clumps can bias 
the spectrum toward lower temperatures. 
This issue can be settled performing a spectral fit analysis as 
in Mathiesen \& Evrard (2001), whose simulations did not include radiative
cooling. 

The projected metallicity profiles are displayed as a function of
distance in panel (b). Data points represents the mean profile from the nine 
cooling
flow clusters of De Grandi \& Molendi (2001). For model A3, the iron profile
of L00A3 is the one in better agreement with data.
The profile of L39A3 also has a shape similar to that of L00A3, but with a 
much higher abundance at $r \sim 0.02 r_{200}$ ( $Fe/H \simeq 0.7-0.8$).
The profile of L40A3 agrees fairly well with data only for the first
three radial bins. The overall shape is similar to that of the other two
runs, but with a lower amplitude. The systematic difference between the profiles
of runs L39A3 and L40A3 is mostly due to the different dynamical histories
of the two clusters. Therefore, there are final uncertainties in the final
profiles which can be as high as $\sim 50 \%$ and are related to the 
cluster dynamical evolution. 
At outer radii all the profiles show a radial decay steeper than that of 
data points, which in fact can be considered to have an almost constant
profile. It appears very difficult to modify the model parameters of the 
simulations in a way such that the simulated profiles match the data 
points at outer radii, without also increasing central abundances.

The flatness of the observed profiles suggest that at, early epochs, a series
of merger events has erased the existing metallicity gradients.
Accordingly, the metal abundances from SNII do not show significant 
spatial gradients, while the iron abundance gradient can be attributed
to SNIa \cite{dub20}. In this scenarios the metal excess distribution 
is an indicator of the optical light distribution of early type galaxy, as it 
has been at least partially confirmed by De Grandi \& Molendi (2001). 
This is not the behavior of the simulated cluster sample. Fig. \ref{VHS}
shows for two clusters the iron radial distribution originated from 
both SNII and Ia.
 The iron distributions are very similar, because the timescales of metal
ejection is much higher for SNIa than for SNII this is indicative that 
the dynamical evolution of the two clusters has been very smooth.
As already stressed, this cluster sample has been chosen for the regularity
of its members. In order to asses in a significant way the effects of
dynamical evolution on the shape of the metallicity profiles it is
necessary to perform a statistical analysis over a large (say $\simgt 40$)
cluster sample. This task is left to a future paper, where a number of 
issues will be investigated with a statistically robust cluster sample.
For the sake of clarity in panel (b) are also shown the profiles of 
the runs L40A and L40A2, to be compared with that of L40A3.
These profiles are clearly below the measured values and confirm the 
parameters of model A3 as the ones yielding the best agreement with data.

The iron profile of L119A3 is inconsistent with the data points; the iron
abundances are smaller at all the radii. The profile is very similar to the 
ones of L40A2 and L40A. 
This is a failure of model A3 that is hard to reconcile within the 
framework of the adopted prescriptions, unless the diffusion of metals in the 
ICM is 
  characterized by a minimum diffusion length.
 The most important difference of 
\La119 is that this is the coldest of the four test clusters, with a 
mass-weighted
temperature of $\sim 1keV$. For this cluster a proper comparison of the
simulated iron profiles with the mean metallicity profile of the sample
is not possible. The averaged profile is that of 9 cooling flow clusters,
with minimum temperatures $\simgt 4 keV$.
Without measured profiles for cool clusters is therefore difficult 
to put observational constraints on different model parameters 
from the simulated profiles.

\begin{figure}
\vfill
\centerline{\mbox{\epsfysize=12.0truecm\epsffile{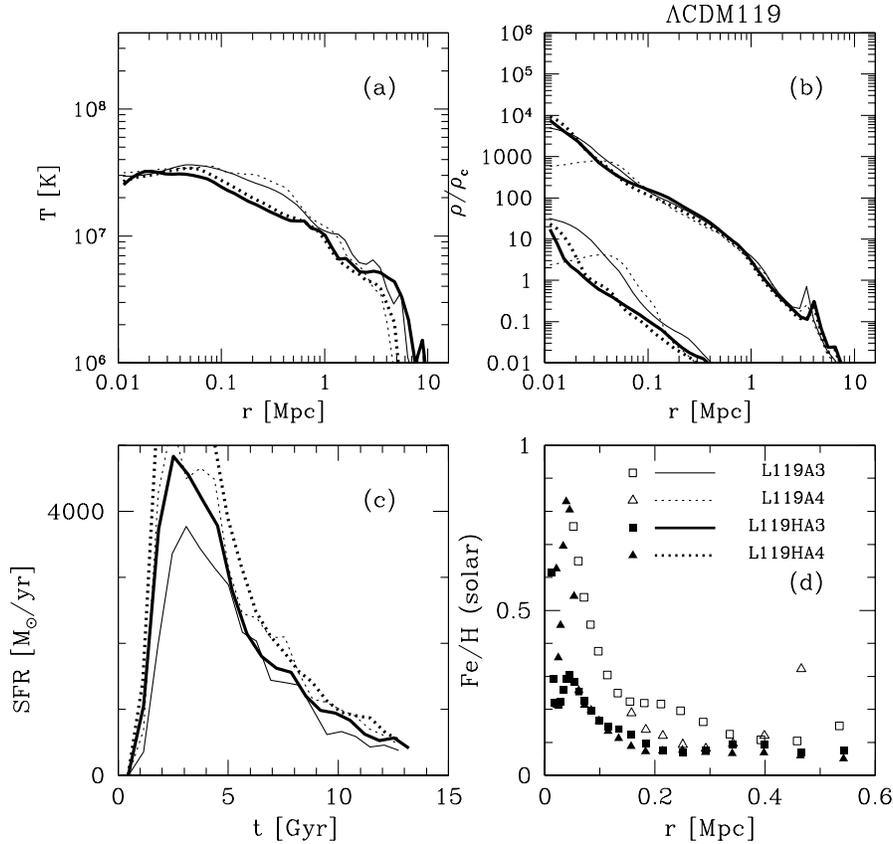}}}
\caption{
The final profiles of the simulations L119A3 and 
L119A4 are compared against those of the corresponding high-resolution runs
(thick lines). {\bf (a)}: radial temperature profiles. {\bf (b)} : Gas density
profiles, the profiles with $\rho(0) / \rho_c \simlt 10^2 $ are those of the 
 gas metallicity. {\bf (c)} : SF rates versus age. {\bf (d)}:
Iron abundance profiles.
}
\label{hsn}
\end{figure}

Global values, $A_{Fe}=M_{Fe}/M_H$, of the iron abundances for the simulated 
clusters are compared in panel (c) for models A and A3 against the 
estimated values for the nearby cluster sample of 
Matsumoto et al. (2000) sample. These values have been taken from column 4 
of Table 1 of the Matsumoto et al. paper and are estimated at a 
radius of $0.5 h^{-1} Mpc$. For model A3 there is a fair agreement with data,
while model A yields values of $A_{Fe}$ which are marginally consistent
with the allowed uncertainties ( $90 \%$ c.l.). The run L119HA3 has a 
value $A_{Fe} \simeq 0.2$, about a factor $\sim$ two smaller 
than that expected extrapolating the sample average below the minimum
 ($\simeq 2keV$ ) cluster sample temperature. 
This is connected with the results shown in the previous panel. 
Without a minimum smoothing length there is 
a clear tendency in the simulations to produce a lesser
amount of iron than that inferred from observations for cool clusters.

The open diamond of Fig. \ref{lx}(c) is the value of $A_{Fe}$ ( $ \sim 0.47$, 
see Table \ref{te}) for the run L119HA5. For this simulation, a minimum
value $h_{min}^s=12 Kpc$ has been assumed for the smoothing length of metals 
in the ICM. Such a high value of $A_{Fe}$ is clearly more indicated 
to account for the iron abundance of low temperature clusters. 
The observational evidence of an iron abundance decreasing with $T_X$ 
is however statistically weak \cite{mu97,fin01}.
Furthermore, an increase of $A_{Fe}$ for cool clusters can be an artifact
due to the presence of a dominant galaxy in the cluster central region.
For example, Fukazawa et al. (2000) have removed the contribution of the
central region for those clusters with a cD galaxy and for their
cluster sample found that there is not a significant correlation $A_{Fe}-kT_X$.
 This is also in agreement with what was found by Finoguenov 
et al. (2001).
A statistical comparison between the $A_{Fe}$ distribution generated by
the simulations and the one from real data has been performed for the models
discussed above. The statistical tests applied to the \Af~ data and the 
\Af~ distribution of the simulated cluster sample are the Student t-test for
 the means, the F-test for the variances and the Kolgomorov-Smirnov (KS)
test for the distributions. The corresponding probabilities give the 
significance level that, according to the analyzed quantities, the two sets
are originated from the same process.
The results are presented in Table \ref{tf}. Model A is clearly ruled out,
model A3 is marginally consistent. This is because the value \Af~of 
L119HA3 is significative to lower the mean of the sample. A much better 
agreement is obtained if the global iron abundance is given by the run L119HA5
for the cluster \La119.

Finally, the final values of the bolometric X-ray luminosity are shown in 
panel (d) as a function of the cluster temperature. Mass-weighted temperature
have been used as unbiased estimators of the spectral temperatures
\cite{ma01}.
The luminosity $L_X$ of the simulated clusters is calculated  as
described in sect. 2. 
Data points are those of Fig. 11 of Tozzi \& Norman
(2001). For the sake of clarity, not all the points of the Figure are plotted 
in the panel. For a consistent comparison with data, a central 
region of size $50 h^{-1} Kpc$ has been excised \cite{mar98} in 
order to remove the contribution to $L_X$ of the cooling flow central
region. For model A3, the $L_X$ of the simulations are in excellent
agreement with data over the entire range of temperatures.
An additional run has been performed for the cluster with the lowest
temperature (\La119). The parameters of this run (A4) are the same of 
model A3, expect for a SN feedback energy
of $10^{50}$ erg being used for both SNII and Ia. This value is $10 \%$ of 
that of
model A3 and has been considered in order to investigate the effects on
final X-ray properties of the amount of heating returned to the ICM
by SNe. As can be seen, for model A4 the final $L_X$  of the run 
L119HA4 is very similar to that of L119HA3. The result demonstrates that 
final X-ray luminosities of the simulations are not sensitive to the 
amount of SN feedback energy that has heated the ICM. In fact, a 
simulation run with a zero SN energy returned to the ICM ($\eps=0$)
yields very similar results. These results are particularly relevant
in connection with the recent proposal \cite{br00} that the X-ray properties
of the ICM are driven by the efficiency of galaxy formation, 
rather than by heating due to non-gravitational processes. 
Final profiles for models A3 and A4 are investigated in more
detail in Fig. \ref{hsn} and \ref{esn}.

\begin{table}
\centering
\begin{minipage}{140mm}
\begin{tabular}{@{}cccc@{}}
data-model & $p_t$ & $p_F$ & $p_{KS}$ \\ 
\hline \hline
$A_{Fe}$-{A} & .003 &  .032 & .001 \\ 
$A_{Fe}$-{A3} & .082 &  .694 & .120 \\ 
$A_{Fe}$-{A3(5)}$^{a}$ & .689 &  .993 & .756 \\ 
$f_{gas}(500)$-{A3} & .02 &  .565 & .018 \\ 
$f_{gas}(200)$-{A3} & .001 &  .221 & .001 \\ 
$f_{gas}(500)$-{A3(19)}$^b$ & .143 & 0.849  & 0.163 \\ 
\hline
\end{tabular}
\caption{ 
Statistical tests applied to distributions obtained from real data and
simulated clusters.
$p_t$ is the Student t--test applied to the means,
$p_F$ is the F--test for variances and $p_{KS}$ is the KS statistic to
discriminate two distributions.
$A_{Fe}$ is the measured iron abundances distribution of Fig.~\ref{lx}, 
 panel (c). $f_{gas}$ refers to the observational sample of the two top panels 
of Fig.~\ref{fgs}. 
The parameters of the model are indicated in Table \ref{td}.
For the simulated cluster sample the numerical resolution of the runs
is L00, L39, L40 and L119H.
$^a$: This sample has L119HA5 in place of L119HA3.
$^b$: This sample has only three clusters, but their initial conditions have 
been set with $\Omega_b h^2=0.019$.
}
\label{tf}
\end{minipage}
\end{table}

\subsection {SN heating of the ICM and cooled gas fractions}
\begin{figure}
\vfill
\centerline{\mbox{\epsfysize=12.0truecm\epsffile{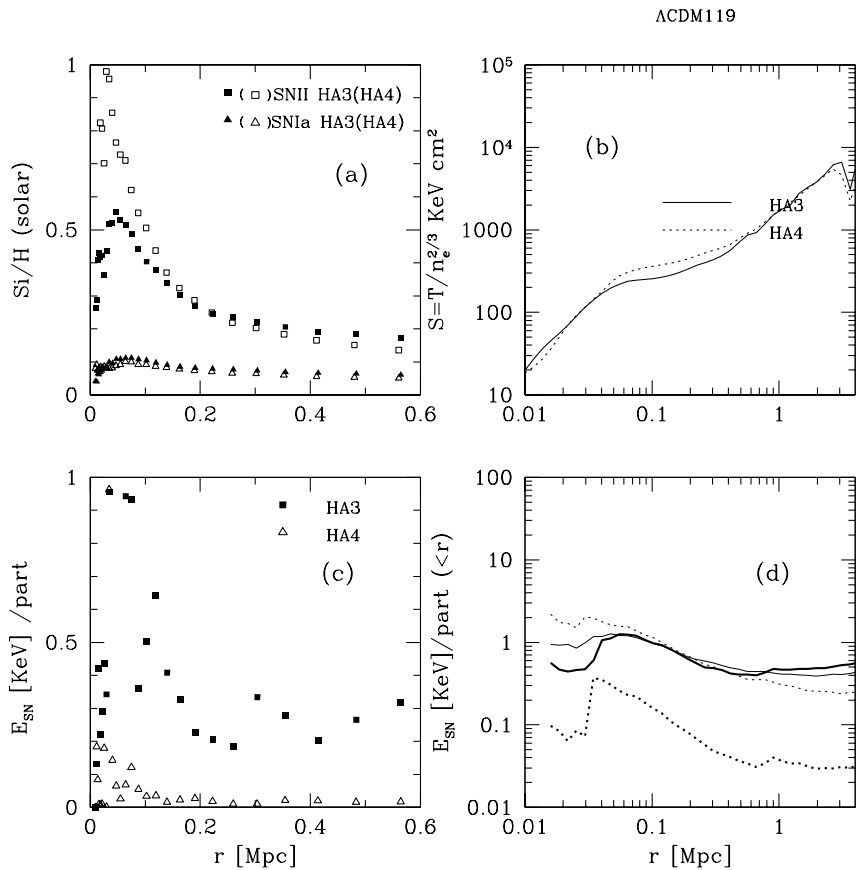}}}
\caption{ 
For the cluster $\Lambda$CDM119 final distributions of the run L119HA3
are compared against those of L119HA4 ($\eps=10^{50} erg$).
Panel {\bf {(a)}} shows the Si abundance of the gas, originated 
from both SNII and SNIa  metal enrichment of the ICM.  
{\bf {(b)}}: Entropy profiles $S=T(r)/n_e(r)^{2/3}$, where 
$n_e=\rho_g(r)/\mu_e m_p$ and $\mu_e \simeq 1.14$.
{\bf{(c)}}: Distribution of the gas particle energy injected by SNe.
{\bf{(d)}}: As in panel {\bf{(c)}}, but for the cumulative distribution
$E_{SN}(<r)/N_p(<r)$, where $N_p=M_g(<r)/\mu m_p$. Thin lines correspond to
$E_{SN}(<r)/N_p(<r)$ as obtained from Si abundance.
}
\label{esn}
\end{figure}

In Fig. \ref{hsn}, the final profiles of the runs L119A3 and L119A4 are 
compared against those of L119HA3 and L119HA4. The main differences are between 
the profiles of the standard resolution run and the high-resolution run.
The profiles of the runs for the models A3 and A4 are very similar,
 with the only important difference between the two models being the  
SFR at early redshifts. The temperature profiles are almost identical,
so energy feedback from SNe is not relevant to determine final gas
properties. This conclusion is also supported from the entropy profiles 
of Fig. \ref{esn}.

In Fig. \ref{esn}, final distributions of the two runs L119HA3 and L119HA4 
are compared. In panel (a) are plotted the Si abundances synthesized 
in the explosions of SNII and SNIa.
There are not large differences between the corresponding profiles.
The Si abundances of SNIa are almost identical. For SNII, the Si
abundances profile of L119HA4 is higher than that of L119HA3 in the cluster
central region ($ \simlt 0.1 Mpc$). These differences follow because the 
run L119HA4 has at early redshifts a higher SFR than that of L119HA3
(see also Fig. \ref{hsn}c). This is a consequence of the reduced thermal 
pressure and higher gas densities in the former simulation. This effect is
not relevant for the SNIa explosions because the involved timescales are 
much higher than those of the early generations of SNII.

The cumulative injected energy per baryon within the radius $r$ is defined as 
the ratio $E_{SN}(<r) \mu m_p /M_g(<r)$, where $E_{SN}(<r)$ is the total SN
energy injected within that radius, $M_g(<r)$ is the corresponding 
gas mass.
This ratio is shown in panel (d) for the two runs. The ratio has a radially 
decreasing behavior, for the run L119HA3 drops from $\sim 1keV/part$ at 
$r \simeq 0.1 Mpc$ to $\simeq 0.5 keV/part$ at $r \simeq 1 Mpc\simeq r_{200}$.
There is a small drop in the cluster inner region, presumably  due to
a biased estimate because of the small number of simulation particles for 
$r \simlt 40 Kpc$. 
For the run L119HA4 the energy per particle ratio has  
a shape similar to that of the corresponding ratio for the run L119HA3,
at $r \sim 1 Mpc$ takes the value of $\sim 0.04 keV/part$, which is 
about $\sim 10 \%$ of the value of the parent run with $\eps=10^{51} erg$.
The binned distribution (panel c) is much more noisy than the cumulative 
one, but it qualitatively confirms the expected behavior.

 In order to assess the amount of heating from the SNe, this ratio
is often inferred from the measured abundance of metals \cite{fin01}.
 Si abundances are particularly useful because of the similar yields 
for different SN types. The ratio estimated from the amount of Si abundance 
$A_{Si}$ is $\eps A_{Si} \mu m_p /y_{Si}$ with $y_{Si}\sim0.12$ being the 
average SN yield of Si. The estimated ratio (thin lines) is in good agreement at all the radii with the values obtained from the simulation.
For $\eps=10^{51}erg$ the average energy per particle of the cluster \La119 is
then $\sim 0.5 keV /part$ at the virial radius ($ r_{200} \sim 1Mpc$). 
This cluster has a virial temperature $\sim 1.5 keV$, from a sample
of 18 relaxed clusters with temperature below $\sim 4 keV$. 
Finoguenov, Arnaud \& David (2001, see Fig. 9) find an average SN injected
energy of $\sim 0.5 keV/part$ for a cluster with a temperature $ \sim 1 keV$.
There is therefore a good agreement between the measured amount of thermal
energy per particle associated with SN feedback and that predicted by the 
simulations with the model parameters of Fig. \ref{esn}.

The gas entropy is defined as $S(r)= k_B T(r)/n_e(r)^{2/3}$, where
$n_e$ is the electron number density. For the two runs L119HA3 and 
L119HA4 the entropy profiles in units of $keV cm^2$ are plotted in panel
(b) of Fig. \ref{esn}.
An important result that follows from the radial  dependence of the two
 profiles is that they are almost identical. A simulation with a zero SN energy
yields a very similar profile. This means that final ICM properties are 
largely unaffected by the amount of energy injected by SNe into the ICM. 
This follows because most of the energy is injected in the cluster
central region where the gas density is higher and cooling is very efficient
to radiate away the energy of the reheated gas.
Energy feedback from SNe can modify the ICM state for small clusters or
groups with $T_X \simlt 1keV$.
Another important result from the entropy profiles of 
Fig. \ref{esn} is the entropy level of the gas at $r \sim 0.1 R_V \simeq 0.1
r_{200} \simeq 0.1 Mpc$. At this radial distance the gas entropy of the 
simulated cluster is $\sim 200 keV cm^2$, a value consistent with a set of
observations \cite{pon99,lly20,wu02a} for a system with $T_X \sim 1.5 keV$.
Together with the good agreement between the $L_X-T_X$ relation obtained from
simulations with that from observations these findings provide strong
support for the radiative cooling model proposed by Bryan (2000).

According to this scenario \cite{br00,vo01,wu02a,mu02,da02} the efficiency of
galaxy formation is higher in groups and cool clusters than in hot clusters.
The low-entropy cooled gas is removed because of galaxy formation, and 
is replaced by the inflow of the surrounding high-entropy gas.
The higher efficiency of galaxy formation for cool systems explains 
the central entropy excess over the self-similar predictions \cite{pon99}.
One of the most important observational consequences of the cooling model
is that the fraction of hot gas $f_g =M_g/M_T$ increases going from cool
clusters to hot clusters. Conversely, the fraction $f_{star}$ of
cooled gas turned into stars should decrease with the mass of the system.

Observational evidence for a dependence of $f_g$ with $T_X$ is controversial
\cite{dav90,da95,mo99,ar99,rou20,ba01}. The main difficulty is the 
extrapolation to virial radii of the X-ray data for cool clusters, which
can bias the estimates of the cooled gas fraction. According to Roussell
et al. (2000) there is not strong observational support for an increase 
of $f_{gas}$ with $T_X$. This is in disagreement with the findings of
David, Jones \& Forman (1995), who show that the gas fraction $f_{gas}$ 
increases with the X-ray temperature. 
According to Mohr, Mathiesen \& Evrard (1999) there is a weak dependence
of $f_{gas}$ with $T_X$ in their X-ray ROSAT sample of 45 galaxy clusters.
A constant $f_{gas}$ is inconsistent with the data at $95 \%$ confidence
level. 

Arnoud \& Evrard (1999) have analyzed the temperature dependency 
of $f_{gas}$ for a sample of 24 clusters with accurate measured 
temperatures and low cooling flows. They have determined $f_{gas}$ for 
two radii enclosing gas overdensities $\delta=500$ and $\delta=200$.
Total masses have been estimated according to two different methods.
The $\beta$-model (BM) assumes an isothermal gas density profile in
hydrostatic equilibrium with the shape determined according to the X-ray
surface brightness. The virial theorem (VT) estimates the cluster total mass
according to virial equilibrium at a fixed density contrast.
The relation is calibrated from a set of numerical experiments \cite{ev96}
and is not sensitive to the assumed cosmological model. 
According to Arnaud \& Evrard (1999) it is difficult to draw conclusions
on the dependence of $f_{gas}$ on the cluster temperature. The results
depend on the chosen model. For the BM model, the difference in the value 
of $f_{gas}$ between subsamples of cool clusters and hot ($> 4keV$) clusters
is not statistically significant. For the VT model this difference is at the
$3\sigma$ level at a radius corresponding to a gas overdensity $\delta=500$.
Arnaud \& Evrard (1999) indicate a conservative upper limit of $\sim 30 \%$ in
the $1\sigma$ fractional error of $f_{gas}$.

\begin{figure}
\vfill
\centerline{\mbox{\epsfysize=12.0truecm\epsffile{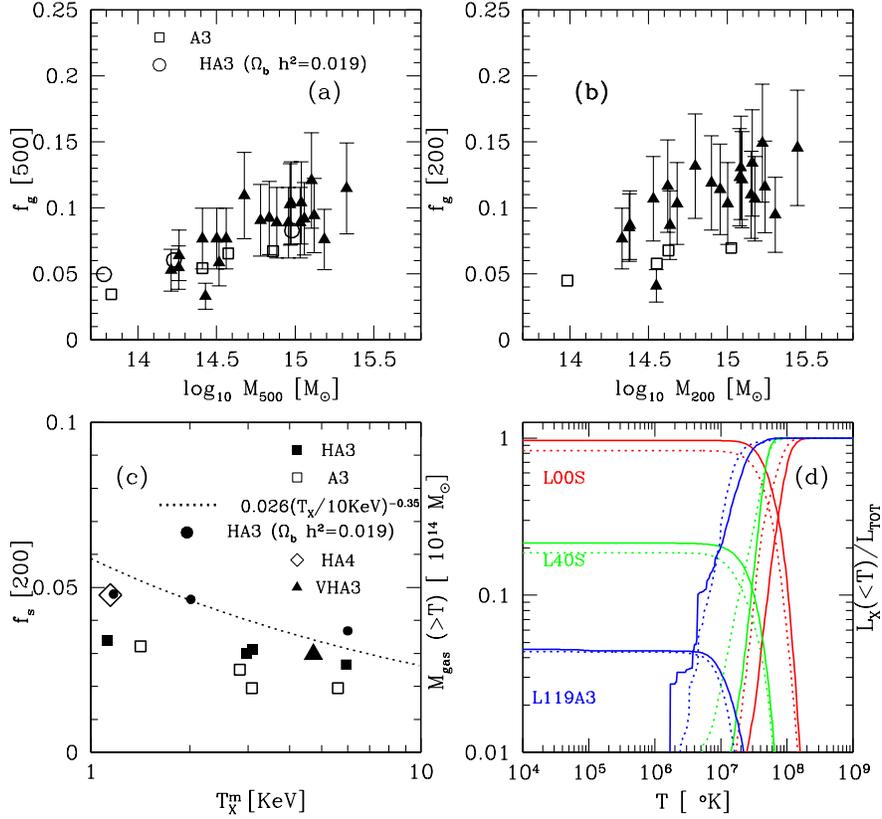}}}
\caption{
Top panels: gas fraction $f_g=M_g(<r)/M_T(<r)$ versus $M_T$ within the radius
within which $\rho/\rho_c=\delta$, {\bf (a)} is for $\delta=500$ and
 {\bf (b)} is for $\delta=200$. Data points from Fig. 3 of Arnaud \& Evrard 
(1999, VT model).
The open circles in panel {\bf (a)} are the values of $f_g$ for runs 
 with $\Omega_b=0.019 h^{-2}$. 
{\bf (c)} : The ratio  of the mass of cold gas turned into stars to the 
total mass is shown as a function of the cluster temperature. The dashed 
line is the Bryan (2000) best-fit $f_{star}$ to data, rescaled by the factor
$f_{baryon}(here)/f_{baryon}(Bryan)=0.1/0.16$.
{\bf (d)}: From the left, gas mass above a given temperature $T$ is plotted as
a function of the temperature itself. Dashed lines are for the very-high 
resolution runs VH, while L119HA3 is for \La119. From the right: the fraction of total X-ray luminosity that is emitted 
from the gas below a given temperature is shown for the runs corresponding 
to the gas masses on the left.
}
\label{fgs}
\end{figure}

For the cluster sample of Arnaud \& Evrard (1999), the gas fraction $f_{gas}$ 
is shown as a function of the total cluster mass in the top panels of
Fig. \ref{fgs}. The sample distributions are derived from the VT
model at a density contrast of $\delta=500$ (top left panel) and $\delta=200$ 
(top right panel). The data points are those in the top panels of Fig. 3 
of Arnaud \& Evrard (1999), with an assumed $\simeq 30 \%$ uncertainty.
(top right panel). For the model A3 the corresponding values of $f_{gas}$ 
at $\delta=500$ and $\delta=200$ from the numerical 
cluster sample are also shown as open squares. There is a clear tendency for 
the two simulated cluster distributions
to follow the $f_{gas}$ distribution of the data, but with a reduced 
amplitude. The result of a statistical comparison are reported in 
Table 6. The $f_{gas}$ distribution of model A3 at $\delta=500$ is 
inconsistent with data with a high significance level ($95 \%$ c.l.),
and the situation at $\delta=200$ is even worse. For this overdensity,
an extrapolation of X-ray data up to the required radius has been 
performed \cite{ar99} in order to evaluate the corresponding data points.
Therefore, possible biases can undermine the estimate of $f_{gas}$ at 
$\delta=200$. For $f_{gas}[\delta=500]$ a possible source of disagreement
between the numerical distribution and that of data points lies in the 
assumed value of the cosmological baryonic density
$\Omega_b$ in the simulations. Here it has been assumed $\Omega_b h^2=0.015$, 
but recent 
measurements \cite{bu98} favor $\Omega_b h^2 =0.019$. In correspondence to
this value of $\Omega_b$ in the initial conditions of the simulations,
the values of $f_{gas}[\delta=500]$ are displayed in Fig. \ref{fgs}(a)
for a subsample of three simulated cluster (\La00, \La119 and a new
cluster with virial mass $\sim 2.5 \cdot 10^{14} h^{-1} \msu$).
The distribution is now in a better agreement with the data, and the confidence 
levels for rejecting the null hypothesis are now below $95\%$. 
Therefore it is fair to say that the ICM gas fractions obtained here are
 consistent with available observational estimates.

The predictions of the radiative cooling model for the ICM evolution
 have been recently investigated in a number 
of papers, either through analytical methods \cite{br00,vo01,wu02a,vo02,wu02b}
 or with numerical simulations \cite{mu01,mu02,da02}.
Results from the employed methods clearly show a decreasing $f_{star}$ with
the cluster temperature.
The dependence of $f_{star}[\delta=200]$ on $T_X$ is shown in Fig. \ref{fgs}(c)
for the four test clusters. Open squares are for model A3  with standard
resolution and filled squares refers to the high-resolution runs.
For comparative purposes the $f_{star}$ predicted by the analytical model
of Bryan (2000) is also shown. Because of the different values of $\Omega_b$ 
and $\Omega_m$ in the paper of Bryan, $f_{star}$ has been approximately 
rescaled 
by the factor $f_b(here)/f_b(Bryan)=0.1/0.16=0.625$. There is a fair agreement
of the model with the simulations at high temperatures, but for $T_X\sim 1keV$
$f_{star}$ of the simulations is below the model predictions by a factor 
$\sim$ two. 
The disagreement at low temperatures is ameliorated for the runs with 
$\Omega_b h^2 =0.019$, in such a case $f_{star}(Bryan)$ must be rescaled 
by the factor $\sim 0.81$. The predicted value of $f_{star}$ is now 
$\sim 0.075$ at $T_X \simeq 1keV$, which is a factor $\simeq 50\%$ higher
than the value $\sim 0.05$ of $f_{star}$ at the same temperature for the runs 
 with $\Omega_b h^2 =0.019$.

The comparison of $f_{star}$ with data is a controversial issue \cite{ba01},
also because of the lack of firm bounds on the allowed uncertainties
\cite{rou20}; however, it is worth stressing that Roussel et al. (2000)
found a weak dependence of the gas fraction on cluster temperature when 
the estimates of virial masses are calibrated from numerical simulations 
rather than inferred from the $\beta$-model. The dependence is not 
statistically significant because of the large size of the sample error bars.
 A comparison of $f_{star}$ predicted by an analytical model has been
performed by Wu \& Xue (2002b) against the Roussell et al. (2000) data.
The theoretical predictions are of about twice as high as  
 the sample values. 
From Fig. \ref{fgs}(c) the values of $f_{star}[\delta=200]$ for the
three runs with $\Omega_b h^2 =0.019$ can be compared with the 
analytical predictions of the Wu \& Xue (2002b) model 
 (cf. Fig. 2 of their paper). The values of $\Omega_b$ and $h$ are 
slightly different, but there is a broad agreement. At $T_X \sim 1keV$
$f_{star}(here) \simeq 0.05$ is lower than the theoretically 
predicted value $\simeq 0.07$. As noted by Wu \& Xue theoretical 
models of cooling yield a dependence of $f_{star}$ on $T_X$ steeper than 
that found in hydrodynamical simulations and observations.
At low temperatures $f_{star}$ of the simulations is in better agreement
with the Rousell at al. (2000) data, but the observed uncertainties
do not allow firm conclusions. The simulations results for $f_{star}$ are also 
in qualitative agreement with those of the radiative model of Muanwong et 
al. (2002). In units of $\Omega_b/\Omega_m$ the values of $f_{star}(f_g)$  
range from $\sim 0.38(0.46)$ from $M_{200}=0.8 \cdot 10^{14} h^{-1} \msu$
to $\sim 0.34 (0.56)$ for a cluster with a virial mass of 
 $M_{200}\sim2.4 \cdot 10^{14} h^{-1} \msu$. These values are not very 
different from those which can be inferred from the distributions plotted 
in Fig. 3(b) on Muanwong et al. (2002) for their radiative model.
The cosmological parameters of the simulations of Muanwong et al. (2002)
are identical to those assumed here ($\Omega_m$ is a $\sim 20\%$ higher),
and the numerical resolution is broadly similar.

Finally, the effects of the numerical resolution on the predicted final amount
of cooled gas are shown in panel (d) of Fig. \ref{fgs}.
As a function of the gas temperature the fractional contribution to 
the total X-ray luminosity is plotted from the right for runs of three clusters
(\La00, \La40 and \La119) with different resolutions. For the first two clusters
the ratio of $L_X(<T)/L_T$ for very high-resolution runs are compared against
that of standard resolution. The ratio $L_X(<T)/L_T$ of the run L119A3 is
instead compared against that of L119HA3.
From the left, the gas masses above a given gas temperature are plotted for 
the corresponding runs. The distribution of the gas mass versus the 
gas temperature shows that resolution effects are most important in the 
low-temperature region of the gas temperature distribution, whereas
X-ray luminosities are largely determined by the high-temperature part
of the distribution. This demonstrates  that the fraction of cold gas 
depends on the numerical resolution, as can be seen from Fig. \ref{fgs}(c),
but is not a source of concern as far as X-ray properties are interested.

For the cluster \La119, the quantity, $M_{gas}(>T)$ is weakly dependent
 on the numerical resolution of the simulations.
From Fig. \ref{fgs}(c), one can see that $f_{star}$ at low temperatures also 
seems to converge
as the resolution is increased. This issue is strictly related to the global
value of the baryonic fraction of cooled gas 
$\Omega_{cold}/\Omega_b=f_b(global)$. Observational
upper limits indicate $f_b(global)\simlt 10\%$ \cite{ba01}. 
According to Balogh et al. (2001), the global fraction $f_b(global)$ is expected to increase as the resolution  limit of the simulation is increased,
 unless a feedback model is incorporated in the simulation.
The results of the high-resolution runs suggest that at low temperatures 
convergence is being achieved for $f_{star}$ and that the feedback model
implemented here effectively regulates the amount of cooled gas.
This is strongly indicated by the large value of $f_{star}$ ($\sim0.05$)
when the SN feedback energy is reduced (open diamond of Fig. \ref{fgs}(c), 
which corresponds to the run L119HA4). 
For clusters with temperatures above $\simgt 2keV$, the values of $f_{star}$
appear to converge for simulations with a number of gas particles 
 $N_g \simgt 200,000$.
This is shown in Fig. \ref{fgs}(c), where for the cluster \La00 the value of
$f_{star}$ for the very-high resolution run L00VHA3 (filled triangle)
 is $\sim 0.03$ and is close to that of $f_{star}$ for L00HA3. Note 
that the mass of the gas 
particle for the run L00VHA3 is only a $\sim10\%$ higher than that of L119HA3 
($\simeq 7\cdot10^8 \msu$).
These issues can be clarified in deeper details with a set 
of cosmological simulations with different resolutions that incorporate
the SN feedback scheme implemented here. Such analysis is beyond the scope of 
this paper and will be considered in a future work.

\section{SUMMARY AND CONCLUSIONS}
 In this paper results from a large set of hydrodynamical SPH simulations
of galaxy clusters have been used to investigate the dependence of iron 
abundances and heating of the ICM on a number of model parameters. The simulations 
have been performed with different numerical resolutions and a numerical
cluster sample covering nearly a decade in cluster mass. The modeling of the 
gas physical processes in the simulations incorporates radiative cooling,
star formation and energy feedback. The gas is metal enriched from SN
ejecta of type II and Ia, and the cooling rate depends on the local
gas metallicity. This allows us to follow self-consistently the ICM evolution 
also
for cool clusters, whose departures of cluster scaling relations from 
self-similarity are most important.

The metal enrichment of the ICM is governed by a number of model parameters.
Theoretical uncertainties on the shape of the IMF and nucleosynthesis stellar 
yields lead to final iron abundances which can differ by a factor $\sim$ two.
These two parameters have then been kept fixed for a large fraction of the
performed simulations. This is done in order to restrict the full range of the 
parameter space. They have been chosen so that the corresponding final
amounts of iron in the ICM are the ones that most closely agree with that
indicated by observations.
The ejected metals are distributed among gas neighbors according to the 
SPH formalism. Final iron abundances have been found to be sensitive to 
the shape of the smoothing kernel of the ejected metals $W_Z$ and to the 
constraints on the corresponding smoothing lengths $h^s$.
For a certain choice of the above parameters, it has been found a fair agreement
between the observed radial metallicity profiles with those obtained from
the simulated clusters. Global iron abundances are also consistent with
measured values.
There are however a number of issues which are still open. 

i) For cool clusters 
the ejection parameters are not well determined because of the lack of measured
iron profiles, also constraints from global iron abundances suffer from 
selection biases in the cluster samples.
The simulated profiles have a radial decay steeper than observed. This
difference could be mainly due to the lack in the simulated cluster sample
of mergers which have efficiently remixed the metal content of the ICM.
The small size of the numerical cluster sample does not allow to reach
firm conclusions. In order to investigate the correlation between the shape of 
the final metallicity profiles with the cluster dynamical evolution it is 
then necessary to analyze simulation results from a statistically 
robust (say $\simgt 40$) cluster sample.

ii) A discrepancy with observations is given by the radial behavior of the 
projected emission-weighted temperature profiles, which have a steep rise
at $r \simlt 0.1r_{200}$. This feature is not shared by observations, for 
the innermost bin the simulated temperatures are higher by a factor $\sim$ two 
than the measured values.
This difference cannot be easily accommodated by the gas physical modeling
of the simulations. Because of radiative cooling, an increase of the 
gas temperature toward the cluster centre is expected as a consequence
of entropy conservation \cite{wu02a}.
A certain amount of cooled gas is present in the core of the clusters and is
responsible of the strong decline of the temperature profiles at
distances very close to the cluster centres ($ r \simlt 0.02 r_{200}$).
This feature could help to reconcile the differences with the shapes 
of the observed profiles.
As found by Mathiesen and Evrard (2001), the measured spectral fit 
temperatures can be biased toward lower values because of line cooling.
An X-ray spectral fit analysis is then necessary to resolve the 
discrepancies between the simulated and measured temperatures profiles in the
cluster core regions.

The ICM X-ray properties of cool clusters are largely unaffected by
the amount of feedback energy injected by SNe, with the luminosity-temperature
relation being in excellent agreement with data. The final gas entropy 
distribution has been found almost independent on the heating efficiency,
for $T_X\sim 1keV$ the core entropy excess is in agreement with estimated 
values.
These findings support the radiative cooling model of Bryan (2000), where 
the ICM X-ray scaling relations are driven by the efficiency of galaxy
formation. The model predicts that at the cluster virial radius the 
fraction of hot gas $f_g$ is positively correlated with the cluster
temperature $T_X$. A number of observational uncertainties prevents to put 
tight constraints, but the $f_g-T_X$ distribution of the simulated cluster
sample is not inconsistent with available estimates from X-ray data.

To explain the steepness of the luminosity-temperature relation an alternative 
to the cooling model is the preheating scenario, where the gas has been 
heated by an energy injection that occurred at early epochs.
Simulation results \cite{bi01,mu02,bo02} show that a number of observed
X-ray properties can be reproduced in the preheating model.
Future X-ray observations of cool clusters and groups
with Chandra will help to discriminate between the two models. 
It is worth noting 
that simulations of galaxy clusters in the preheating scenario 
must also include radiative cooling and SN feedbacks.
Therefore, the agreement with observations of X-ray properties reproduced
here by the simulated clusters in the cooling model suggests that additional
heat sources, if required, are most likely to affect the ICM properties of
systems with temperatures below $\simlt 1keV$.

\section*{Acknowledgments}

The author is grateful to M. Arnaud, S. De Grandi and P. Tozzi for 
providing their data files. G. Carraro and S. Recchi are also 
gratefully acknowledged for clarifying comments on chemical
evolution. The author would like to thank also an anonymous 
referee for useful comments and suggestions that improved the presentation of
 the paper.

\clearpage

\end{document}